\pdfoutput=1
\documentclass[fleqn,usenatbib]{mnras}

\usepackage{newtxtext,newtxmath}

\usepackage[T1]{fontenc}

\DeclareRobustCommand{\VAN}[3]{#2}
\let\VANthebibliography\thebibliography
\def\thebibliography{\DeclareRobustCommand{\VAN}[3]{##3}\VANthebibliography}

\usepackage{graphicx}	
\usepackage{amsmath}	
\usepackage{mathtools}
\usepackage{float}
\usepackage{threeparttable}

\newcommand{\msun}{$M_{\odot}$}

\newcommand{\mj}{$M_{J}$}
\newcommand{\sigmagas}{$\Sigma_\text{gas}$}
\newcommand{\sigmadust}{$\Sigma_\text{dust}$}
\newcommand{\deltadust}{$\delta_\text{dust}$}
\newcommand{\deltagas}{$\delta_\text{gas}$}
\newcommand{\gasdust}{$\Delta_\text{g/d}$}
\newcommand{\rsub}{$R_\text{sub}$}
\newcommand{\rgap}{$R_\text{gap}$}
\newcommand{\rcav}{$R_\text{cav}$}
\newcommand{\shgas}{S/H$_\text{gas}$}
\newcommand{\ohgas}{O/H$_\text{gas}$}
\newcommand{\chgas}{C/H$_\text{gas}$}

\newcommand{\sostack}{SO $7_7-6_6$ + $7_8-6_7$ }

\title[Volatile sulfur in HD~100546]{Spatially resolving the volatile sulfur abundance in the HD~100546 protoplanetary disk}

\author[L. Keyte et al.]{
Luke Keyte,$^{1}$\thanks{E-mail: luke.keyte.18@ucl.ac.uk}
Mihkel Kama,$^{1, 2}$
Ko-Ju Chuang,$^{3}$
L. Ilsedore Cleeves,$^{4}$
Maria N. Drozdovskaya,$^{5}$
\newauthor Kenji Furuya,$^{6}$
Jonathan Rawlings,$^{1}$
Oliver Shorttle$^{7}$
\\
$^{1}$Department of Physics and Astronomy, University College London, Gower Street, WC1E 6BT London, United Kingdom \\
$^{2}$Tartu Observatory, Observatooriumi 1, T\~{o}ravere 61602, Tartu, Estonia\\
$^{3}$Laboratory for Astrophysics, Leiden Observatory, Leiden University, P.O. Box 9513, NL-2300 RA Leiden, the Netherlands.\\
$^{4}$Department of Astronomy, University of Virginia, Charlottesville, VA 22904, USA\\
$^{5}$Center for Space and Habitability, Universität Bern, Gesellschaftsstrasse 6, CH-3012 Bern, Switzerland\\
$^{6}$National Astronomical Observatory of Japan, Osawa 2-21-1, Mitaka, Tokyo 181-8588, Japan\\
$^{7}$Department of Earth Sciences \& Institute of Astronomy, University of Cambridge, United Kingdom\\
}

\date{Accepted on 20th December 2023}

\pubyear{2023}

\begin{document}
\label{firstpage}
\pagerange{\pageref{firstpage}--\pageref{lastpage}}

\maketitle

\begin{abstract}
Volatile elements play a crucial role in the formation of planetary systems. Their abundance and distribution in protoplanetary disks provide vital insights into the connection between formation processes and the atmospheric composition of individual planets. Sulfur, being one of the most abundant elements in planet-forming environments, is of great significance, and now observable in exoplanets with JWST. However, planetary formation models currently lack vital knowledge regarding sulfur chemistry in protoplanetary disks. Developing a deeper understanding of the major volatile sulfur carriers in disks is essential to building models that can meaningfully predict planetary atmospheric composition, and reconstruct planetary formation pathways. In this work, we combine archival observations with new data from ALMA and APEX, covering a range of sulfur-bearing species/isotopologs. We interpret this data using the \textsc{dali} thermo-chemical code, for which our model is highly refined and disk-specific. We find that volatile sulfur is heavily depleted from the cosmic value by a factor of $\sim 1000$, with a disk-averaged abundance of S/H $\sim10^{-8}$. We show that the gas-phase sulfur abundance varies radially by $\gtrsim 3$ orders of magnitude, with the highest abundances inside the inner dust ring and coincident with the outer dust ring at $r\sim 150$ to $230$ au. Extracting chemical abundances from our models, we find OCS, H$_2$CS, and CS to be the dominant molecular carriers in the gas phase. We also infer the presence of a substantial OCS ice reservoir. We relate our results to the potential atmospheric composition of planets in HD~100546, and the wider exoplanet population.
\end{abstract}





\begin{table*}
\caption{Observational parameters for sulfur-bearing species in HD 100546 from ALMA and APEX data. Disk-integrated flux upper limits are at 3$\sigma$ level, denoted by `<'.}
\begin{threeparttable}
\label{table:observations}      
\centering 
\begin{tabular}{l l l l l l l l l}     
\hline\hline       
                      
Molecule & Transition & $\nu$ (GHz) & $\Delta_{\nu}$ (GHz) & Channels & E$_\textrm{up}$ (K) \tnote{$\dagger$} & $A_\textrm{ul}$ (s$^{-1}$) \tnote{$\dagger$}& Beam Size & Flux (W/m$^2$)\\ 
\hline
\hline
\multicolumn{9}{c}{ALMA 12m array} \\
\hline
   SiS               & 5-4                      & 90.769    & 0.117  & 960 & 90.8 & $1.19 \times 10^{-5}$ & 0.10"$\times$0.07" & < 3.28$\times 10^{-19}$\\  
   $^{13}$C$^{34}$S  & 2-1                      & 90.923    & 0.059  & 480 & 6.5 & $1.34 \times 10^{-5}$ & 0.10"$\times$0.07" & < 1.84$\times 10^{-19}$\\
   $^{13}$CS         & 2-1                      & 92.491    & 0.059  & 480 & 6.7 & $1.41 \times 10^{-5}$ & 0.10"$\times$0.07" & < 4.85$\times 10^{-20}$\\
   SO$_2$            & 3$_{1,3}$-2$_{0,2}$      & 104.029   & 0.117  & 480 & 7.7 & $1.01 \times 10^{-5}$ & 0.09"$\times$0.06" & < 9.71$\times 10^{-20}$\\
   SO$_2$            & 10$_{1,9}$-10$_{0,10}$   & 104.239   & 0.117  & 480 & 54.7 & $1.12 \times 10^{-5}$ & 0.09"$\times$0.06" & < 9.93$\times 10^{-20}$\\
   $^{34}$SO$_2$     & 10$_{1,9}$-10$_{0,10}$   & 104.392   & 0.117  & 480 & 54.6 & $1.10 \times 10^{-5}$ & 0.09"$\times$0.06" & < 1.00$\times 10^{-18}$\\
   $^{30}$SiS        & 6-5                      & 105.059   & 0.117  & 480 & 17.7 & $1.88 \times 10^{-5}$ & 0.09"$\times$0.06" & < 1.88$\times 10^{-18}$\\
   Si$^{34}$S        & 6-5                      & 105.942   & 0.234  & 960 & 17.8 & $1.90 \times 10^{-5}$ & 0.09"$\times$0.06" & < 1.36$\times 10^{-19}$\\
\hline
\multicolumn{9}{c}{APEX} \\
\hline
   o-H$_2$S          & 1$_{1,0}$-1$_{0,1}$      & 168.763   & 0.090   & 160 &  27.9 & $2.65 \times 10^{-5}$  & 37.1"$\times$37.1" & < 1.06$\times 10^{-21}$ \\
\hline                  
\end{tabular}

\begin{tablenotes}\footnotesize
\item[\tnote{$\dagger$}] Line frequencies, upper energy levels (E$_\text{up}$), and Einstein A coefficients (A$_\text{ul}$) are taken from the Cologne Database for Molecular Spectroscopy \citep[CDMS; ][]{CDMS_muller_2001, muller_2005, endres_2016} and the Leiden Atomic and Molecular Database \citep[LAMDA; ][]{LAMDA_schoier_2005}.
\end{tablenotes}

\end{threeparttable}
\end{table*}

\section{Introduction}

Sulfur is one of the most abundant elements in the Universe, essential to astrochemistry and biological processes. Understanding sulfur's path from the interstellar medium, through disks, and into planets is of increasing importance, relevant to the study of planetary formation and the search for life. Despite this, sulfur chemistry in star- and planet-forming environments is poorly understood. In the interstellar medium the total sulfur abundance is S/H = $1.32 \times 10^{-5}$, of which 99\% is in the gas phase \citep{savage_sembach_1996, asplund_2009}. In molecular clouds, however, a factor of $\geq$10 depletion is observed \citep[e.g.][]{tieftrunk_1994}. This missing sulfur constitutes one of the major outstanding questions in astrochemistry. Similarly, the main carrier of volatile sulfur in protoplanetary disks is also unknown. Chemical models predict the dominant volatile carriers to include species such as H$_2$S, OCS, SO, SO$_2$ and CS, but past observations have typically lacked the sensitivity to definitively address the volatile sulfur budget in disks \citep[e.g.][]{wakelam_2004}. In some systems, upper limits determined from non-detections have been used to provide useful constraints \citep[eg.][]{maps_12_legal2021}.

Some clues about sulfur's chemical pathway through molecular clouds and into planetary systems come from observations of our own Solar System, where sulfur-bearing species are routinely detected in comets. These are most often in the form of H$_2$S and S$_2$ \citep{mumma_charnley_2011}, but also include a large variety of other species. Observations of 67P/Churyumov-Gerasimneko, for example, uncovered a wealth of sulfurated species in the cometary coma, including H$_2$S, S, S$_2$, S$_3$, S$_4$, SO, SO$_2$, H$_2$CS, CS$_2$, OCS, C$_2$H$_6$S, CH$_3$SH, and NH$_4^+$SH$^-$ \citep{le_roy_2015, calmonte_2016, altwegg_2022}. Such observations provide valuable insights into the conditions and processes that existed during the formation of the Solar System. \looseness=-1

By comparison, observations of sulfur-bearing species in protoplanetary disks are relatively rare, with only six different molecules having been detected thus far; CS, SO, H$_2$S, SO$_2$, H$_2$CS, and SiS \citep[e.g.][]{dutrey_2011, guilloteau_2013, phuong_2018, legal_2019, Booth_2021, law_2023}. Modelling of the sulfur budget in protoplanetary disks suggests that sulfur is highly depleted from the gas phase, but the split between rocky, icy, and gas-phase components is not well understood. Previous studies predict that the main refractory component is iron sulfide, FeS \citep{pasek_2005}, and that conversion from volatile molecules to FeS likely proceeds in the early stages of disk evolution when gas-phase molecules react with warm, iron-rich dust grains \citep{lauretta_1996}. \citet{Kama19} inferred the refractory sulfur reservoir in a sample of disks around Herbig Ae/Be stars to be $89\pm 8\%$ of the total sulfur, leaving $11\pm 8\%$ distributed to volatile molecules. These stars have shallow surface convective zones, in which newly accreted material dominates the photospheric composition. Signatures of sulfur-bearing species in the photospheres of such stars are indicative of material passing through the inner cavity of the disk, and these systems therefore provide a unique laboratory for studying volatile sulfur budgets \citep{jermyn_kama_2018}.

In this work, we present observations from the ALMA 12m array and APEX telescope, covering nine transitions of eight sulfur-bearing species and their isotopologues in the protoplanetary disk HD~100546. We use the \textsc{dali} 2D physical-chemical disk modelling code to constrain the abundance and spatial distribution of volatile sulfur-bearing species, and determine the total elemental gas-phase sulfur abundance. The aim is to gain a deeper understanding of the major volatile sulfur carriers in protoplanetary disks, and provide insights as to how their spatial distribution is linked to exoplanet bulk and atmospheric composition.


\section{Observations} \label{observations}

\subsection{Target: HD 100546}

HD~100546 is a well-studied 2.49 $\pm 0.02$ \msun \ Herbig Be star of spectral type B9V, with an estimated age of $\sim$ 5 Myr \citep{Arun_2019}, at a distance of 110$\pm$1 pc \citep{GAIACollaborationetal2018}. The star hosts a bright transition disk with an inclination of 44$^{\circ}$ and position angle of 146$^{\circ}$ (e.g. \citealt{Walsh_2014, fedele_2021}), with a total observable dust mass of $\sim 10^{-3}$ \msun \ (\citealt{wright_2015, miley_2019}). Micron-sized grains are observed out to at least 1000 au \citep{ardila_2007}. ALMA band 7 observations show that mm-sized grains are abundant out to $\sim$ 230 au, but are depleted inside of $\sim13$ au and again from 40 to 150 au \citep{Walsh_2014, fedele_2021}. The inner cavity hosts a smaller inner disk extending from $\sim 0.25$ to $1$ au containing $3\times 10^{-10}$ \msun \ of dust \citep{Benisty_2010}. Modelling suggests that the size of the inner cavity and outer gap are consistent with radial zones cleared of material by two planetary companions: a 20\mj\ planet at $\sim 13$ au and a 15\mj \ planet at $\sim 55$ au \citep{Walsh_2014, quanz_2015, Currie_2015, pinilla_2015}. This interpretation is supported by infrarad imaging with the Gemini Planet Imager, where point sources have been resolved at the same radial locations as the proposed planetary candidates \citep{Currie_2015}. There is also evidence for a warped or misaligned inner gas disk, traced by deviations from Keplerian rotation in the CO J=3-2 emission. \citep{Walsh2017, Booth2018}.

CO isotopologue observations suggest that the molecular gas extends to at least $\sim 390$ au \citep{Walsh_2014}. A variety of  atomic and molecular lines have been observed, including lines of atomic oxygen and atomic carbon \citep{Thi2011, Kama2016a, Kama2016b}, spatially resolved CO 3-2 emission \citep{Walsh_2014}, various CO isotopologues, and three H$_2$O lines \citep{Bruderer2012, meeus2012, meeus_2013, Kama2016b}. Spatially resolved observations of SO with ALMA reveal emission from inside the dust cavity, likely related to thermal desorption of ices at the cavity wall \citep{booth_2023a}. This SO emission has also been linked to ongoing planet formation, thought to be tracing shocked gas in the vicinity of a circumplanetary disk. CS has also been detected, making HD\;100546 one of only three disks where both SO and CS have been detected together, the others being HD 169142 \citep{law_2023, booth_2023b} and AB Aur \citep{riviere_marichalar_2023}. Azimuthal asymmetries observed in the emission morphology and spectral lines of each of these species have previously been used to infer an azimuthal C/O variation in the disk \citep{keyte_2023}. No other sulfur-bearing species have been detected in HD~100546 thus far. However, \citet{Kama19} measured the total sulfur abundance in the stellar photosphere to be log(S/H) = $-5.37 \pm 0.5$, thus providing an initial constraint on the volatile sulfur content of accreting material.

\begin{figure}
\centering
\includegraphics[clip=,width=1.0\linewidth]{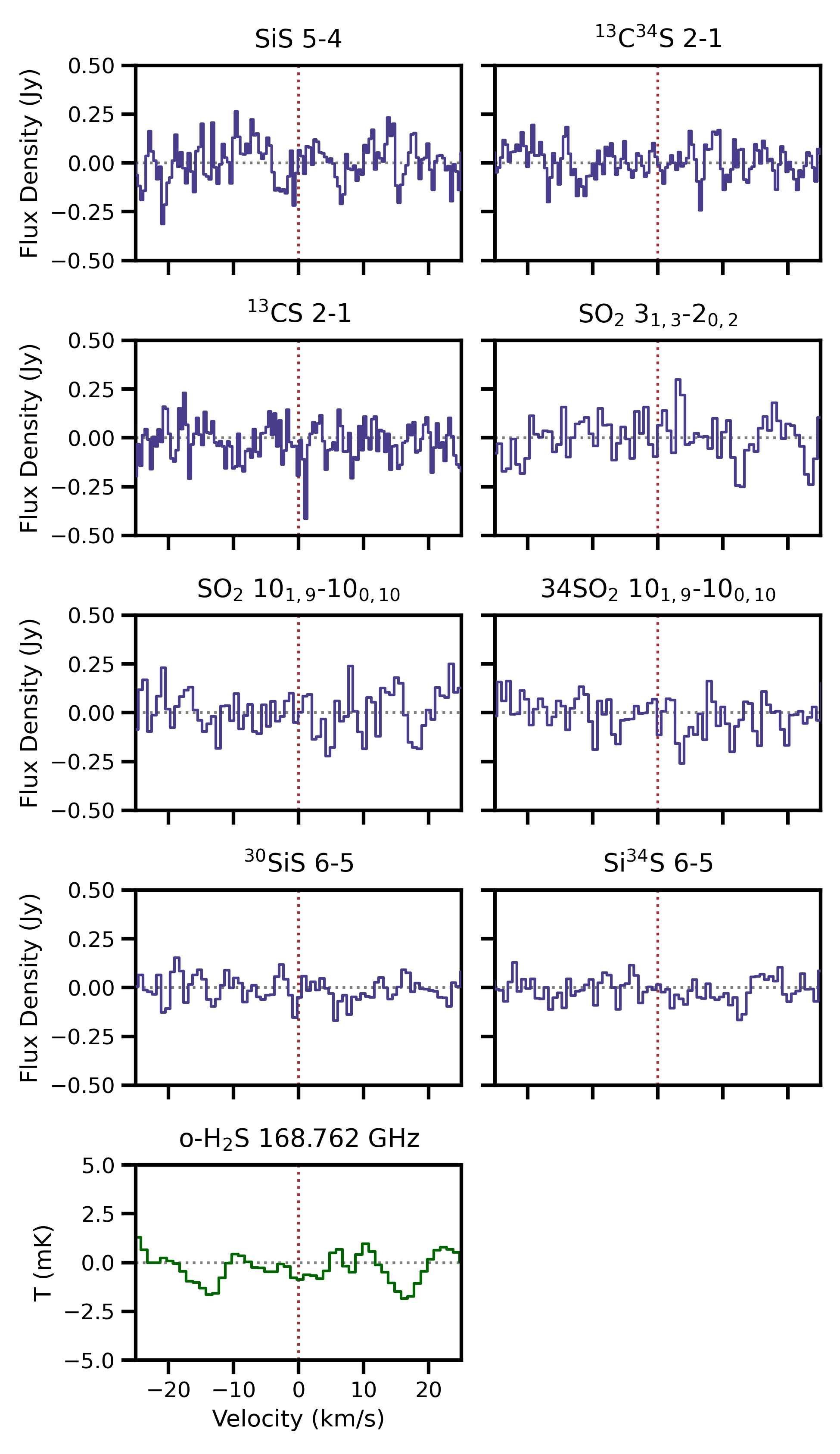}
\caption{Spectra obtained from ALMA (blue) and APEX (green) for each of the transitions listed in Table 1. No spectral lines are detected. All spectra have been adjusted
for the system velocity $V_\textrm{LSRK} = 5.7$ km s$^{-1}$.}
\label{fig_spectral_lines}
\end{figure}

\subsection{ALMA observations}

We present ALMA observations of HD 100546 in Band 3 from the Cycle 5 program 2017.1.00885.S (PI: M. Kama). Observations were completed in one execution block on October 24th 2017, covering baseline lengths ranging from 41.4 to 13399.7m. The total on-source time was 18.18 minutes. System temperatures varied from 45.4-81.3 K and the average precipitable water vapour was 0.7 mm. J0904-5735 was used as both the bandpass and flux calibrator, while J1145-6954 was used as the phase calibrator. The observations cover nine rotational transitions of eight sulfur-bearing species, outlined in Table \ref{table:observations}. \looseness=-1

The data reduction was completed using the ALMA Pipeline in the Common Astronomy Software Package (CASA) version 5.6.1-8 \citep{mcmullin_2007}. Self-calibration was performed but found to have marginal impact due to low signal-to-noise of the data. Continuum and line imaging were performed with the CLEAN algorithm using Briggs weighting (robust=0.5). Continuum subtraction was performed with the CASA task \emph{uvcontsub}, using a single-order polynomial fit to the line free channels. The spectral resolution and synthesized beam size for each of the transitions are listed in Table \ref{table:observations}. 

The inner dust ring was well-resolved in the continuum data, and the small inner dust disk ($r\lesssim1$ au) was detected. The continuum image is presented in Appendix \ref{appendix:continuum}. No spectral lines were detected in any of the ALMA 12m array data. We initially attempted to extract the spectral lines from the CLEAN image cubes using an elliptical aperture with a 3.9" radius centred on the source. Integrated intensity maps were created from the CLEAN cubes, which yielded no detections. We then employed a number of other techniques in order to increase the signal-to-noise ratio.

First, we stacked the lines in the image plane, by adding together the integrated intensity maps. This resulted in no detections. Next, we applied a Keplerian mask to each of the CLEAN image cubes in order to maximise the signal-to-noise in the image plane \citep{teague2020}. The Keplerian masked cubes were then used to create integrated emission maps, which again failed to result in any detections. Finally, we applied a matched filter to the visibility data, to maximize the signal-to-noise in the \emph{uv}-plane \citep{Loomis_2018}. The matched filter technique utilises a template image cube that samples the \emph{uv}-space in order to obtain a set of template visibilities. These can then be used as a filter, which is cross-correlated with the data in an attempt to detect any weak emission. We created template emission profiles for each of the spectral windows in the ALMA 12m array data by modelling the spectral line emission with the \textsc{dali} thermo-chemical disk modelling code (see section \ref{modelling}). The matched filter was then run for each of the spectral lines individually, but again failed to result in any detections. 

We derived 3$\sigma$ upper limits for the disk-integrated flux for each of the non-detected spectral lines, calculated from the elliptically masked integrated intensity maps (Table \ref{table:observations}), where the mask extended 3.9" (approximately covering the gas disk as traced by CO \citep{Walsh_2014}). Spectra are presented in Figure \ref{fig_spectral_lines}.

\subsection{APEX observations}

Observations of HD\;100546 were carried out in Band\,5 (${\sim}180$ GHz) with the single-pixel SEPIA receiver \citep{belitsky_2018} on APEX, during several runs from March 2021 to September 2021 (program 0108.F-9320, PI: L. Keyte). The observations targeted the o-H$_2$S 1$_{1,0}$-1$_{0,1}$ transition at 168.762GHz. The backend used was FFTS1, with a highest resolution channel space of 0.06 MHz or 0.11 km s$^{-1}$. Observational parameters are listed in Table \ref{table:observations}.

A total of 209 scans were completed, with exposure times ranging from $\sim 0.5$ to $7.5$ minutes. Total observing time was 13.2 hours, with 143.1 minutes on-source. The column of precipitable water vapour was typically 2.5 mm, but ranged from 1.0 to 5.0 mm. Baseline subtraction and other high level reductions were performed with GILDAS/CLASS \citep{pety_2018}, and the data was smoothed to a velocity resolution of $\delta v=1$ km s$^{-1}$.

The o-H$_2$S 1$_{1,0}$-1$_{0,1}$ transition at 168.76 GHz was not detected with APEX. At a spectral resolution of 1 km s$^{-1}$, the achieved rms is 1.53 mK, from which we derived a 3$\sigma$ upper limit on the disk-integrate flux of $1.06 \times 10^{-21}$ W m$^{-2}$. The spectrum is shown in Figure \ref{fig_spectral_lines}.

\subsection{Complementary data}
We also make use of a variety of archival data. These include ALMA ACA observations of sulfur-bearing species first presented in \citet{keyte_2023} (SO, $^{34}$SO, SO$_2$, CS, C$^{36}$S, HCS$^+$, H$_2$CS) and detections of SO 7$_7$-6$_6$ (301.286 GHz) and 7$_8$-6$_7$ (304.078 GHz) first presented in \citet{booth_2023a}.


\section{Modelling} \label{modelling}

Our goal is to use the observational data outlined in the previous section to constrain the abundance of volatile sulfur-bearing species in HD 100546, and determine the radial profile of the total gas-phase elemental sulfur abundance \shgas.

We ran source-specific models using the 2D physical-chemical code \textsc{dali} \citep{Bruderer2012, Bruderer2013}. The code begins with a parameterised gas and dust density distribution and an input stellar spectrum, then uses Monte Carlo radiative transfer to determine the UV radiation field and dust temperature. This provides an initial guess for the the gas temperature, which begins an iterative process in which the gas-grain chemistry is solved time-dependantly. We let the chemistry evolve to 5 Myr to match the best estimate for the age of the disk. Finally, the raytracing module is used to obtain spectral image cubes, line profiles, and disk-integrated line fluxes.

The models presented in this paper are based on the HD\;100546 disk model from \citet{keyte_2023}, with some minor adjustments to improve the fit. See that work for for a full description of the fitting procedure. Here, we briefly restate the key features of the model. A description of the updates and fiducial model parameters are given in Appendix \ref{appendix:model_params_updates}.

\subsection{Disk parameters}
The disk structure is fully parameterised, with a surface density that follows the standard form of a power law with an exponential taper:
\begin{equation}
    \Sigma_\text{gas} = \Sigma_\text{c} \cdot \bigg(\frac{r}{R_c} \bigg)^{-\gamma} \exp \bigg[- \bigg(\frac{r}{R_c} \bigg)^{2-\gamma} \bigg]
\end{equation}
where $r$ is the radius, $\gamma$ is the surface density exponent, $\Sigma_\text{c}$ is some critical surface density, and $R_\text{c}$ is some critical radius, such that the surface density at $R_\text{c}$ is $\Sigma_\text{c}/e$. The scale height is then given by:
\begin{equation}
    h(r) = h_c\bigg(\frac{r}{R_c}\bigg)^\psi
\end{equation}
where $h_\text{c}$ is the scale height at $R_\text{c}$, and the power law index of the scale height, $\psi$, describes the flaring of the disk.

\sigmagas \ and \sigmadust \ extend from the dust sublimation radius (\rsub, defined as the inner edge of the model domain) to the edge of the disk ($R_\text{out}$), and can be varied independantly inside the cavity radius \rcav \ with the multiplication factors \deltagas \ and \deltadust.

The gas-to-dust ratio is denoted \gasdust. Dust settling is implemented by considering two different populations of grains: small grains (0.005-1 $\mu$m) and large grains (0.005-1 mm). The vertical density structure of the dust is such that large grains are settled towards the midplane, prescribed by the settling parameter $\chi$:

\begin{equation}
    \rho_\text{dust, large} = \frac{f \Sigma_\text{dust}}{\sqrt{2\pi}r \chi h} \exp \bigg[ -\frac{1}{2} \bigg( \frac{\pi/2 - \theta}{ \chi h} \bigg) ^{2} \bigg]
\end{equation}

\begin{equation}
    \rho_\text{dust, small} = \frac{(1-f)\Sigma_\text{dust}}{\sqrt{2\pi}rh} \exp \bigg[ -\frac{1}{2} \bigg( \frac{\pi/2 - \theta}{h} \bigg) ^{2} \bigg]
\end{equation}
where $f$ is the mass fraction of large grains and $\theta$ is the opening angle from the midplane as viewed from the central star.

\subsection{Stellar parameters}
The stellar spectrum was modelled by \citet{Bruderer2012} using dereddened FUSE (Far Ultraviolet Spectroscopic Explorer) and IUE (International Ultraviolet Explorer) observations at UV wavelengths, and then extended to longer wavelengths using the B9V template of \citet{Pickles_1998}. The stellar luminosity is $36 \; L_\odot$ \citep{Kama2016b}. The X-ray spectrum was characterized as a thermal spectrum with a temperature of $7 \times 10^7$ K over the 1 to 100 keV energy range, with an X-ray luminosity of $L_X = 7.94 \times 10^{28}$ erg s$^{-1}$.

\subsection{Chemical network}
The chemical network used in our model is based on a subset of the UMIST 06 \citep{woodall2007} network. It consists of 131 species (including neutral and charged PAHs) and 1721 individual reactions. The code includes H$_2$ formation on dust, freeze-out, thermal desorption, hydrogenation, gas-phase reactions, photodissociation and -ionization, X-ray induced processes, cosmic-ray induced reactions, PAH/small grain charge exchange/hydrogenation, and reactions with vibrationally excited H$_2$. Non-thermal desorption is only included for a small number of species (CO, CO$_2$, H$_2$O, CH$_4$, NH$_3$). For grain surface chemistry, only hydrogenation of simple species is considered (C, CH, CH$_2$, CH$_3$, N, NH, NH$_2$, O, OH, S, and HS). The details of these processes are described more fully in \citet{Bruderer2012}. Of particular relevance to this study, the network includes reactions for 38 sulfur-bearing species, including all those listed in Table \ref{table:observations}. The various isotopologs are not included as separate species, but abundances are computed using scaling relations based on typical ISM isotope ratios \citep{wilson_rood_1994}. Initial elemental abundances are listed in \ref{table:initial_abundances}.

\begin{figure*}
\centering
\includegraphics[width=\textwidth]{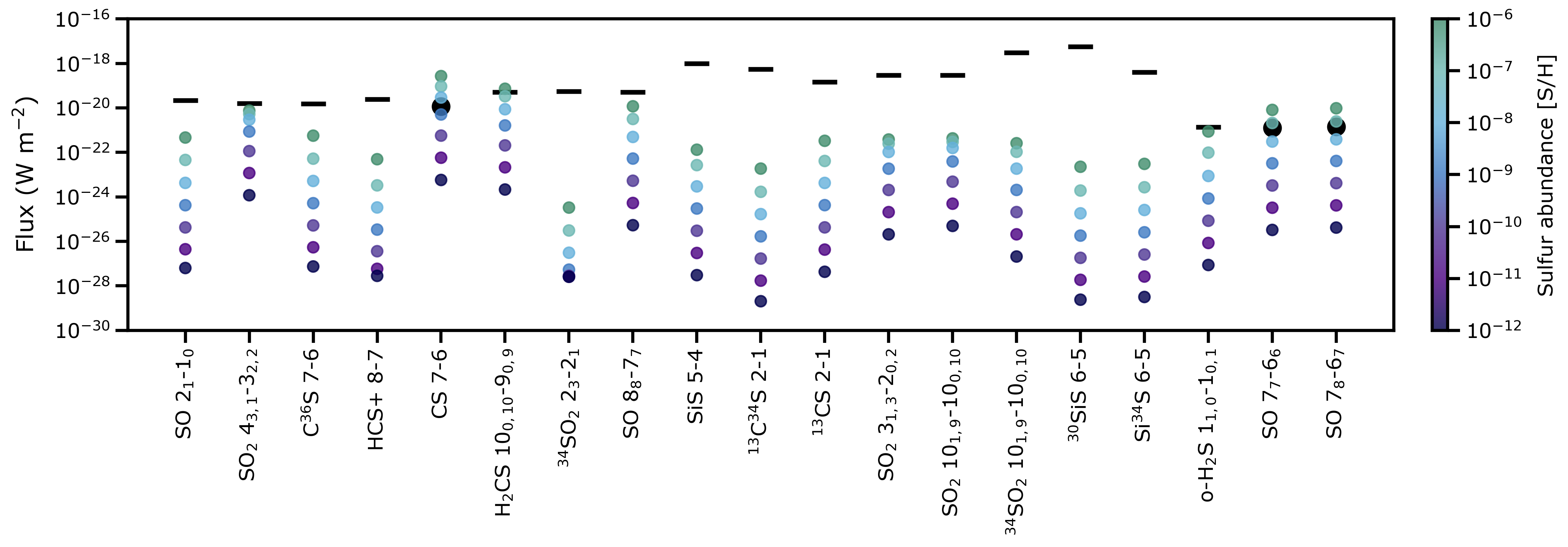}
\caption{Disk-integrated fluxes for each of the transitions listed in Table \ref{table:observations}. Observations are shown in black (where a circle represents a detection and a horizontal line is an upper limit). Model values are shown as coloured circles, where colors denote the initial sulfur abundance used in a given model.}
\label{fig_di_fluxes}
\end{figure*}

\section{Modelling Results} \label{modelling_results}

In order to model emission from sulfur-bearing species in HD 100546, we consider two broad scenarios. Firstly, a best-fit disk-averaged sulfur abundance, and secondly, a sulfur abundance profile which includes both radial and azimuthal variations.

\subsection{Best-fit disk-averaged sulfur abundance} \label{subsec: di_sulfur_abundance}

As step one, we assume a constant volatile sulfur abundance across the entire disk. It is initialised as entirely gas-phase (\shgas), though the code allows species formed in the gas to freeze out if local conditions mandate it. A constant global elemental abundance mimics the approach previously employed in prior analyses of \chgas\ and \ohgas\ in the HD~100546 disk \citep{Bruderer2012, Kama2016b}. We ran models covering a wide range of sulfur abundances S/H from $10^{-12}$ to $10^{-6}$. All other disk parameters remain fixed.

Figure \ref{fig_di_fluxes} shows the disk-integrated fluxes extracted from our models, alongside the observational detections and upper limits derived in section \ref{observations}. Volatile sulfur is found to be severely depleted in the HD\;100546 disk, with the CS detection indicating a best-fit value of $\sim 5 \times 10^{-9}$ and the SO detections indicating a best-fit value of $\sim 5 \times 10^{-8}$. This apparent contradiction is discussed in the next section. The upper limits of the remaining transitions provide no further constraints. These results suggest that the total amount of sulfur in the gas phase is depleted from the canonical cosmic value by at least two orders of magnitude. \looseness=-1

\begin{figure*}
\centering
\includegraphics[width=\textwidth]{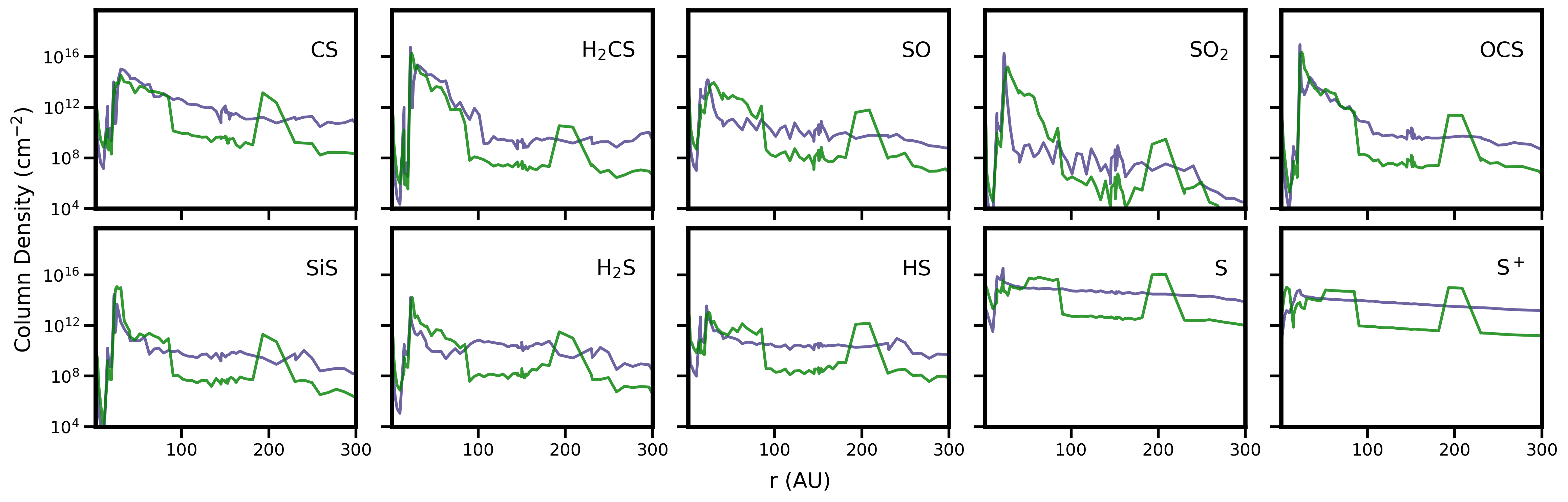}
\caption{Column densities for key sulfur-bearing species, extracted from our models. Purple lines show results for our best-fit model using a single global sulfur abundance (S/H $=10^{-8}$). Green lines show results for our model using a radially varying sulfur abundance profile.}
\label{fig_column_density}
\end{figure*}

Taking S/H $=10^{-8}$ as a fiducial value, we extracted column densities for ten key sulfur-bearing species: CS, H$_2$CS, SO, SO$_2$, OCS, SiS, H$_2$S, HS, atomic S, and S$^+$, shown in Figure \ref{fig_column_density} (purple lines). All of the molecular species display similar morphological radial variations. Column densities peak within the inner dust ring at $\sim 22$ au, then fall off in the outer disk, with varied slopes. The atomic S and S$^+$ column densities vary much more smoothly, remaining consistently high ($\gtrsim 10^{14}$ cm$^{-2}$) well into the outer disk. We present 2D abundance maps for each of these species, in both the gas- and ice-phase, in Appendix \ref{appendix:abundance_maps}.

\subsection{Radially varying sulfur abundance} \label{subsec: radial_sulfur_abundance}

The models outlined in the previous section provide an estimation of the volatile sulfur abundance across the entire disk. However, fitting to disk-integrated line fluxes does not allow us to place constraints on variations in the sulfur abundance at small spatial scales. HD~100546 is known to be highly structured, featuring a small inner dust disk ($r \lesssim 1$ au), two broad millimeter dust rings, and at least two protoplanet candidates. We may then reasonably expect radial variations in the disk chemistry, linked to processes such as radial drift, dust trapping, complex grain surface chemistry, and non-thermal desorption of sulfur-bearing species, which are not explicitly treated in the \textsc{dali} modelling code. Furthermore, we have also previously shown that the asymmetric emission morphology and spectral line profiles of both the CS 7-6 and \sostack observations can be reproduced by a model that incorporates an \emph{azimuthal} variation in the C/O ratio \citep{keyte_2023}. The model from that work uses C/O\;=\;0.5 throughout most of the disk, except for a $60^\circ$ radial arc on the western side of the disk, where the C/O ratio is elevated ($=1.5$). These complexities, which are not accounted for in our `disk-averaged sulfur abundance' model, may explain why there is a discrepancy in the best-fitting value for S/H obtained by matching the CS and SO detections.

In this section, we use the SO $7_7-6_6$, SO $7_8-6_7$, and CS 7-6 observations to undertake a more detailed analysis of the sulfur abundance profile, fitting for both radial and azimuthal variations. We begin by deprojecting the SO $7_7-6_6$, SO $7_8-6_7$, and CS 7-6 integrated intensity maps, in order to extract radial intensity profiles. For each species, we extract a profile that is azimuthally averaged across the entire extent of the disk, ignoring the small region in which the C/O ratio is locally elevated. To begin with, we focus only on fitting these profiles, treating the region where C/O is elevated separately later. The profiles obtained from the two SO transitions are stacked to form a single profile.

\begin{figure*}
\centering
\includegraphics[clip=,width=1\linewidth]{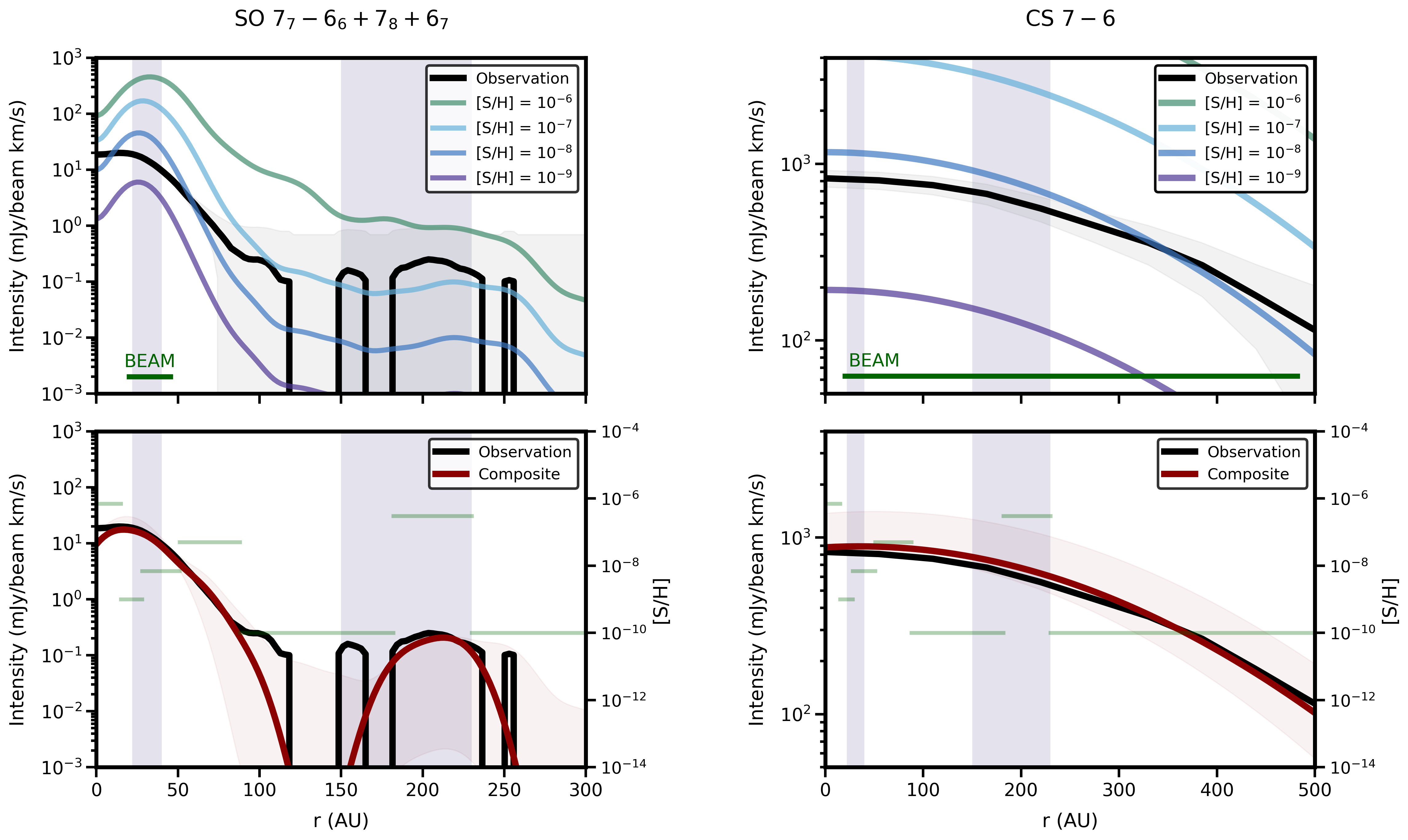}
\caption{\emph{Top row: }Radial intensity profiles extracted from deprojected integrated intensity maps, azimuthally averaged across the entire disk except in the region where C/O is locally elevated. Observations (black, with uncertainties shaded in grey) and models using a range a sulfur abundances (coloured). The semi-major axis of the beam is denoted by the horizontal bar in the lower part of the panel. Shaded blue columns denote the locations of prominent millimeter dust rings. \emph{Bottom row: }Radial intensity profiles extracted from our composite model (red). Horizontal green lines signify the radial regions in which models were spliced together, arranged to match the corresponding sulfur abundance on the secondary y-axis.  Shaded red regions denote the space covered by models computed using the $\pm 1 \sigma$ intervals from the posterior distributions.}
\label{fig_radial_co0p5}
\end{figure*}

Figure \ref{fig_radial_co0p5} (top row) shows the SO and CS radial intensity profiles extracted from the observations, alongside profiles extracted from our models using a range of initial sulfur abundances S/H from $10^{-6}$ to $10^{-9}$. The models have been convolved with a beam of the same size as the observations for a better comparison. The observed SO radial profile peaks at the edge of the inner dust ring, and gradually decreases before sharply falling off at $\sim$125 au. A second region of high abundance is observed in the outer disk between $\sim 150$ to $250$ au, roughly coincident with the outer millimeter dust ring. The observed CS profile peaks at the location of the host star, and falls off smoothly to the outer disk. None of our models are able to simultaneously reproduce the SO or CS emission morphology. Models which reproduce the SO emission in the inner disk fail to reproduce the significant emission peak in the outer region, while models which include significant flux in the outer disk overpredict emission inside of $r \sim 150$ au.

We constructed a composite model by dividing the radial intensity profiles into a predefined number of radial regions, fitting each region with a model that has a sulfur abundance S/H from $10^{-12}$ to $10^{-6}$. In order to efficiently sample the parameter space, we adopted a Markov Chain Monte Carlo (MCMC) approach, making use of the Python package \texttt{emcee} \citep{emcee}. The radial locations at which models are spliced together are treated as free parameters, with uniform priors ranging between 0 and 300 au. The models applied to each region are free parameters drawn from a discretized set of 20 models covering the allowed range of initial sulfur abundances. The SO and CS profiles are fit simultaneously using the same set of free parameters.

In each MCMC iteration, models are first spliced together, then convolved with a beam to match the observation. The goodness-of-fit of the convolved profile is determined using the log-likelihood function:

\begin{equation}
    L = - \frac{1}{2}  \sum \bigg{(}\frac{y_\text{obs}-y_\text{model}}{\sigma}\bigg{)}^2
\end{equation}
where $y_\text{obs}$ are the observational data points, $y_\text{model}$ are the model data points, and $\sigma$ are the uncertainties on the observations.

We ran the MCMC multiple times, each time using a different number of predefined radial regions. We found that eight regions were sufficient to describe variations in the intensity profiles. Using more than eight regions resulted in at least two neighbouring regions sharing the same initial sulfur abundance. In each case, we ran the MCMC for 40,000 steps (excluding 1000 burn-in steps), using 30 walkers. The best-fitting composite model is shown in Figure \ref{fig_radial_co0p5} (lower panel). Horizontal green lines denote the radial extent of each model used to form the composite profile, with the vertical position of each line denoting the sulfur abundance used in that particular model (secondary y-axis).The shaded red area denotes the space covered by models computed using the $\pm1 \sigma$ intervals from the posterior distributions.

The composite model indicates that the volatile sulfur abundance varies by at least $\sim 3$ orders of magnitude throughout different radial regions of the disk. Within the central cavity, the gas-phase sulfur abundance is found to be relatively high (S/H $\sim 7 \times 10^{-7}$), then falls off dramatically inside the inner dust ring (S/H $\sim 1 \times 10^{-9}$). Immediately outside of the dust ring, another region of relatively high abundance is identified (S/H $\sim 5 \times 10^{-8}$), which we note is broadly coincides with the proposed location of the protoplanet HD\;100546 b \citep[]{quanz_2015, Currie_2015}. We discuss this further in section \ref{discussion}. A third region of relatively high abundance (S/H $\sim 3 \times 10^{-7}$) is identified between 180-230 au, broadly coincident with the outer dust ring. This is bounded on both sides by regions of low abundance (S/H $\sim 10^{-10}$), corresponding to regions where large dust grains are highly depleted. The best-fit model appears to be missing SO emission in the region between $\sim 100-125$ au. To investigate this, we manually spliced together a model using a higher sulfur abundance in these regions, matching the SO intensity profile by eye. While this led to a better-fitting SO profile, it caused the CS radial profile to be significantly overpredicted. Possibile explanations for this discrepancy are discussed in section \ref{discussion}.

Having determined the radial gas-phase sulfur abundance profile throughout most of the disk, we now check for consistency by ensuring that our composite model also fits the small region in which the gas-phase C/O ratio is locally elevated. Following the procedure outlined in \citet{keyte_2023}, we compute the abundances within this region, finding the composite model to be a reasonable match to the data (Appendix \ref{appendix: MCMC}). In this region, the observed SO emission tapers off dramitically at $\sim 75$ au, compared to $\sim 120$ in the rest of the disk. The modelling SO emission also tapers of more quickly in this regions, although not to the same degree, cutting off abruptly at $\sim 100$. In the outer disk, the modelled emission is slightly more extended than the observed emission outside of the second dust ring. The modelled CS emission matches the observation extremely well, reproducing the elevated flux in the outer disk. Since both regions can be well-fit using the same radially-varying sulfur abundance profile, we focus only on the large region of the disk where C/O\;=\;0.5 for the remainder of the study, unless otherwise stated.

\begin{figure*}
\centering
\includegraphics[clip=,width=1.0\linewidth]{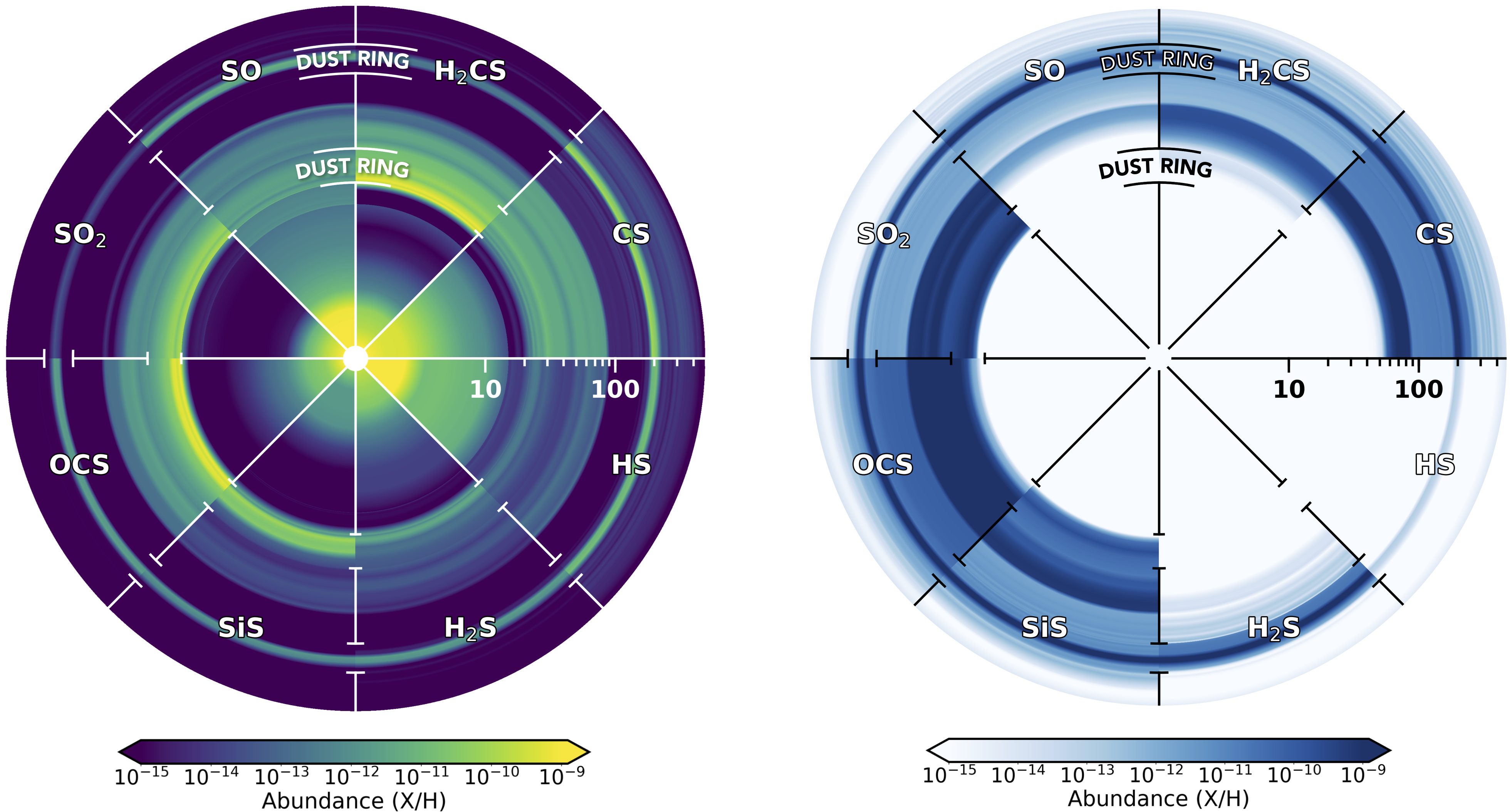}
\caption{Vertically integrated molecular abundances extracted from our composite model, as a function of radius. Gas-phase (left) and ice-phase (right).}
\label{fig_radial_sweep}
\end{figure*}

We present a side-by-side comparison of the modelled molecular abundances extracted from our composite model in Figure \ref{fig_radial_sweep}, for each of the gas- and ice-phase sulfur bearing species. Individual 2D abundance maps are presented in Appendix \ref{appendix:abundance_maps}.

\section{Discussion} \label{discussion}

\subsection{Total elemental gas-phase sulfur abundance}
\label{discussion:1_total_S/H}

We have shown that volatile sulfur is highly depleted in the HD~100546 protoplanetary disk, regardless of whether the abundance is determined as a single global value or as a radially varying gas-phase elemental S/H ratio, which is possible using the full information content of the available data.

Using models with a single sulfur abundance across the entire disk, we find S/H $\lesssim 5 \times 10^{-8}$, over two orders of magnitude lower than the canonical cosmic value. This level of depletion is consistent with the `highly depleted' S-abundance values used in many previous studies. Modelling of the disks around MWC 480, LkCa15, and DM Tau by \citet{dutrey_2011} , for example, showed that an initial abundance value of S/H = $8\times10^{-9}$ was required to reproduce the CS and SO column densities. Similarly, a study by \citet{legal_2019} showed that the observed CS column densities in MWC 480 and LkCa15 could be reproduced with a model using an initial value of S/H $=8\times 10^{-8}$. In both studies, models were not able to fully reproduce the inferred abundances of all sulfur-bearing species, leading to an overprediction of the H$_2$S abundance \citep{dutrey_2011} or underprediction of the H$_2$CS column density \citep{legal_2019}. Such discrepancies are thought to be related to missing reaction pathways, rather that an undepleted sulfur reservoir. Laboratory studies have shown that H$_2$CS can form as an intermediate product via a reaction with H$_2$S and atomic C, for example \citep{galland_2001}. More studies are needed to develop an improved understanding off sulfur chemistry and more complex chemical networks.\looseness=-1

\begin{figure*}
\centering
\includegraphics[clip=,width=1.0\linewidth]{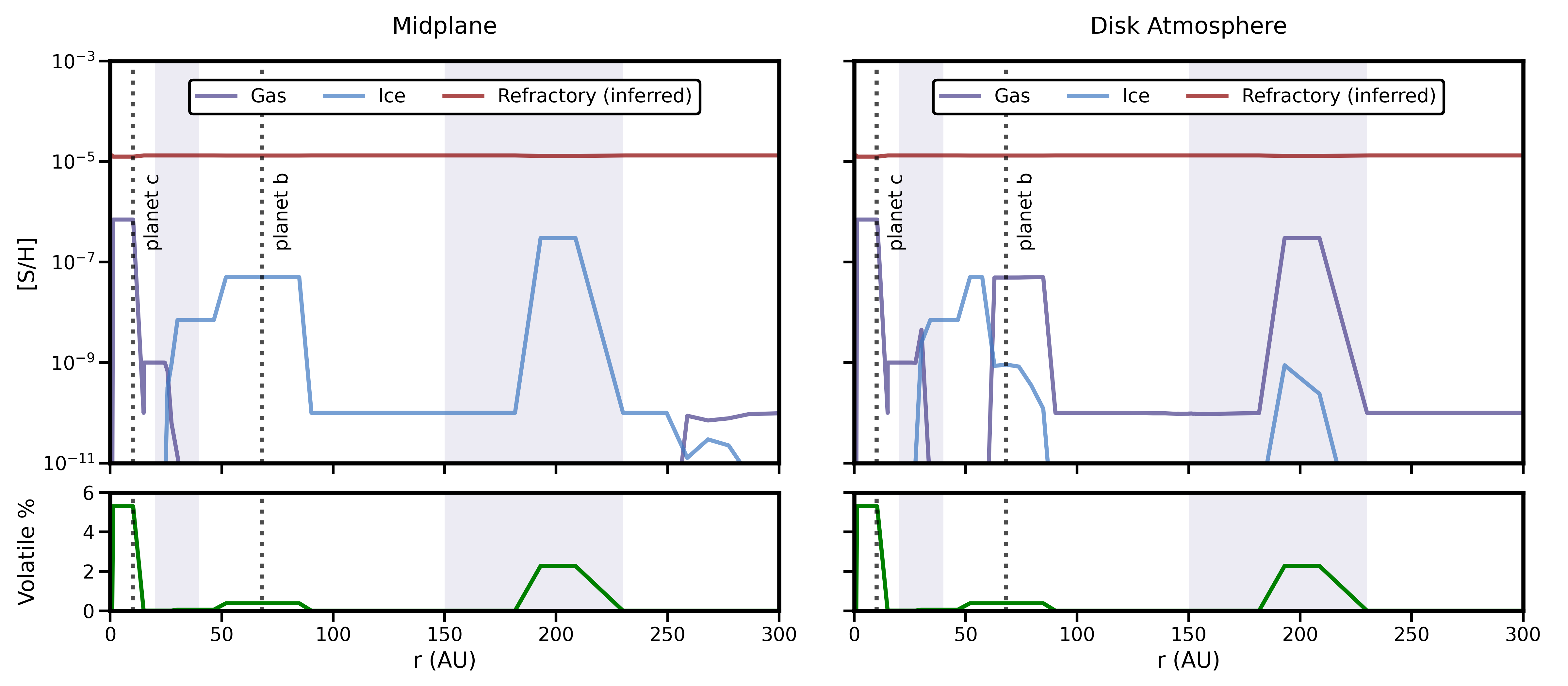}
\caption{Total elemental sulfur abundance of refractories, ices, and gas, as a function of radius. The gas and ice phase abundances are extracted from our composite model, which uses a radially varying sulfur abundance (section \ref{subsec: radial_sulfur_abundance}). The abundance of refractory sulfur-bearing species is inferred from the difference between the cosmic abundance and the total volatile abundance (ice+gas) extracted from the model. Shaded columns represent the location of millimeter dust rings. Dotted black lines denote the location of planetary candidates. \emph{Left: }Midplane abundances. \emph{Right: }Abundances in the warm molecular layer of the disk atmosphere between $z/r=0.2-0.3$. The lower panels in each column show the total amount of volatile sulfur (gas+ice) expressed as a percentage of total sulfur.}
\label{fig_sulfur_budget}
\end{figure*}

The precise impact of missing reaction pathways on modelled abundances is difficult to predict. One such result may be related to our finding that a single sulfur abundance across the entire disk does not adequately reproduce the \sostack and CS 7-6 radial intensity profiles. To better understand why the sulfur abundance may be subject to radial variations, we consider the fraction of the total sulfur budget split between refractories, ices, and gas. Using the radial abundance profile derived in section \ref{subsec: radial_sulfur_abundance}, we plot the proportion of sulfur held in each of these three reservoirs as a function of radius, at both the midplane (Figure \ref{fig_sulfur_budget}, left) and in the warm molecular layer of the disk atmosphere (Figure \ref{fig_sulfur_budget}, right). Values for gas and ice are the total sum of all sulfur-bearing gas and ice species extracted from our model. The abundance of refractory sulfur-bearing species is inferred from the difference between the cosmic abundance and the total volatile abundance (ice+gas) extracted from the model.

In the disk midplane, almost all volatile sulfur is locked up into ices beyond $r \sim 30$ au. The inside edge of the inner dust ring can be considered a transition region, with all volatile sulfur returning to the gas phase within the inner cavity ($r\lesssim22$ au). In the disk atmosphere, the transition from gas to ice at the cavity edge is still apparent, but another transition region is evident in the outer disk at $r \sim 60$ au, where the majority of the volatile sulfur returns to the gas.

The correlation between phase transitions/abundance peaks and the location of prominent millimeter dust rings is suggestive of an underlying connection with grain chemistry or ice adsorption/desorption in both the inner and outer disk. \citet{booth_2023a} propose that SO emission from the inner disk may be partly related to thermal desorption of sulfur-rich ices at the edge of the inner dust cavity, a scenario which is supported by the sharp increase in gas-phase sulfur at the cavity boundary in our model. The same study also suggested that this emission may be due to shocked gas emanating from a circumplanetary disk around a newly-forming planet. This scenario is supported by the fact that the peak in the gas-phase sulfur abundance is located inside the cavity, rather that coincident with the dust ring itself. If this is the case, the sulfur abundance in the inner cavity derived from our fit to the SO radial intensity profile may not a reliable indicator of the abundance of other sulfur-bearing species in this region. Alternatively, the high gas-phase sulfur abundance within the cavity may be linked to the inward drift and sublimation of sulfur-bearing icy grains, which is not incorporated in our dynamically static model.

In the outer dust ring, temperatures are too low for thermal desorption, and the increase in gas-phase volatile sulfur must be driven by some other mechanism. One possible explanation for this is non-thermal desorption from icy grains \citep{booth_2023a}, a scenario which is consistent with rings of H$_2$CO emission observed in the outer disk in a number of other sources \citep[]{riviere_marichalar_2021}. Our models do not include non-thermal desorption of any sulfur-bearing species; if a significant fraction of ices \emph{are} being desorbed in the outer disk, then a lower volatile sulfur abundance than that inferred from our MCMC fitting could reproduce the gas-phase abundance, with a smaller fraction of the total volatiles held in ices.

Taking all radial variations into account, we find that the \emph{total} amount of sulfur locked into volatiles accounts for $\sim$ 0.01 to 5.3\% of the cosmic abundance, dependent on radius. This is broadly in agreement with upper limits on refractory sulfur described in \citet{Kama19} (<93\% of total sulfur), but suggests that an even higher proportion of sulfur is locked into refractories throughout large regions of the disk. The abundance of volatile sulfur in the disk on scales we are sensitive to is also significantly below the sulfur abundance in material currently accreting onto the central star \citep[the total S/H in the accreting material is $\sim 3$ times below solar;][]{Kama2016b, Kama19}. This is consistent with a substantial fraction being transported through the large inner cavity either in small grains or as a semi-refractory reservoir which evaporates at the ${\gtrsim}100-200\,$K temperatures of the main dust ring. \looseness=-1

\begin{figure*}
\centering
\includegraphics[clip=,width=0.9\linewidth]{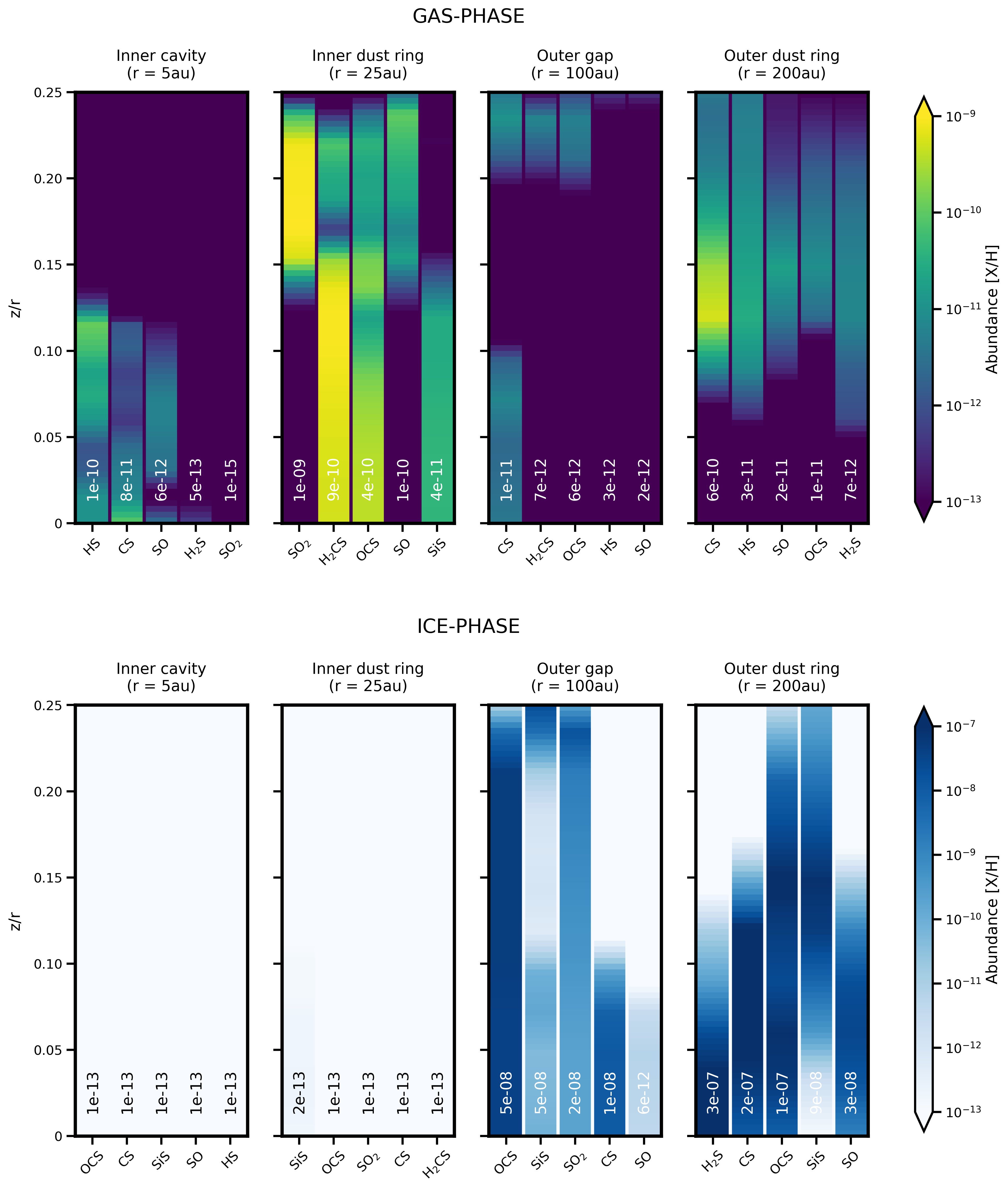}
\caption{Vertical abundance profiles for the main molecular sulfur carriers extracted from our fiducial model, at four fixed radial locations ($r=5, 25, 100, 200$ au). Only the five most abundant species in each region are shown. In each panel, species are ordered by peak abundance (X/H), with the peak value annotated in the respective column. \emph{Top panel: }Gas-phase abundances. \emph{Bottom panel: }Ice-phase abundances.}
\label{fig_vertical}
\end{figure*}

\subsection{Main volatile sulfur carriers}
\label{discussion:2_main_carriers}

In this section, we use abundances obtained from our chemical model to determine the major volatile sulfur carriers in HD~100546 at small spatial scales. Astrochemical models have long predicted dominant carriers in disks to include H$_2$S, SO, SO$_2$, and CS \citep[e.g.][]{pasek_2005}. Observations of comets show that H$_2$ is indeed a major volatile sulfur carrier \citep{calmonte_2016}, which is thought to be due to its preferential formation via hydrogenation of atomic sulfur on grain surfaces \citep{charnley_1997}:
\begin{gather}
    \text{H + S} \rightarrow \text{HS}
    \label{eq:hydrogenation1}\\
    \text{HS + H} \rightarrow \text{H}_2 \text{S}
    \label{eq:hydrogenation2}
\end{gather}
while atomic H can abstract H from H$_2$S:
\begin{gather}
    \text{H}_2 \text{S + H} \rightarrow \text{HS + H}_2
    \label{eq:hydrogenation3}
\end{gather}
and then subsequently reform H$_2$S via reaction \ref{eq:hydrogenation2}. Approximately 4\% of the H$_2$S produced in this way will be released into the gas phase via chemical desorption \citep{oba_2018, furuya_2022, santos_2023}.

Indeed, H$_2$S has the highest inferred abundance of any \emph{ice-phase} sulfur carrier in star-forming regions, although still a factor of $\geq 10$ lower than the cosmic value \citep{van_dishoeck_1998}. Frozen-out H$_2$S can be released from grain surfaces by photodesorption in regions dominated by high UV radiation, although additional photochemistry can lead to destruction via dissociation \citep{jimenez_escobar_2011}. High solid-phase H$_2$S abundances do not therefore necessarily imply a large gas-phase reservoir. In recent years, this has become apparent in observations, with chemical models routinely overpredicting inferred abundances. \citet{phuong_2018}, for example, used the chemical code \texttt{Nautilus} to model H$_2$S detected in GG Tau, overpredicting the inferred column density by a factor of $\sim 10$. In that case, however, the detection itself is thought to be related to the high disk mass rather than low levels of sulfur depletion. Similarly, a study of H$_2$S detected in AB Aur suggests that $99.99$\% of H$_2$S is locked into grain surfaces, and only $\sim 9.5$\% of total sulfur is available for gas-phase chemistry \citep{riviere_marichalar_2021}.

Among other sulfur-bearing molecular species, CS and H$_2$CS are the most widely detected in disks \citep[]{vanderplas2014, Teague_2016, Semenov_2018, legal_2019, maps_12_legal2021}. Conversely, the oxygen-bearing species SO and SO$_2$ have only been detected in a handful of sources \citep[]{guilloteau_2016, fuente_2017, Booth_2021}. Such observations are consistent with the observed overall depletion of volatile oxygen and an elevated C/O ratio in many disks \citep[]{miotello_2019, Zhangetal2020, maps_7_bosman2021}. Typically, inferred abundances of CS and H$_2$CS account for only a small fraction of total sulfur (<1\%), indicating large refractory or unknown volatile reservoir.

It is insightful to compare the outputs of our model to molecular abundances predicted by previous studies. To aid the discussion, we focus on comparing the abundances as a function of height, at four fixed radial locations in the disk (Figure \ref{fig_vertical}). Each radial location is chosen to be representative of particular regions where we might expect large variations in the chemistry: the inner dust cavity ($r=5$ au), inner dust ring ($r=25$ au), outer gap ($r=100$ au), and outer dust ring ($r=200$ au). For clarity, we only display the five most abundant species in each region. Note that in the upper disk atmosphere, all volatile sulfur is in the form of atomic S and/or S$^+$. The precise location where atomic sulfur becomes ionised varies as a function of radius, but is typically at $z/r \sim 0.10$ in the inner dust cavity, and $z/r \sim 0.25$ in the outer disk. The rest of this discussion therefore focuses on the volatile carriers at the midplane and in the warm molecular layer only.

Within the inner cavity, a small molecular reservoir exists close to the midplane, predominantly as HS, CS, and SO, although these account for less that 0.1\% of the total volatile sulfur budget. Gas-phase molecular carriers become much more abundant within the inner dust ring between $r\sim 22-40$ au. At the midplane, almost all volatile sulfur is locked into gaseous H$_2$CS or OCS ($\sim 55$\% and $40$\% respectively). While H$_2$CS is now routinely observed in disks, OCS has yet to be detected and future observations will be needed to assess its potential as a primary gas-phase sulfur reservoir. Further up in the disk atmosphere, OCS and H$_2$CS abundances fall off sharply, with SO$_2$ taking over as the main S-carrier, accounting for $\sim 90$\% of total S volatiles between $z/r \sim 0.15-0.25$.

In the outer dust gap at $r=100$ au, dust temperatures reach $\sim 40$ K in the midplane, and $>99.9$\% of volatile sulfur is frozen out on to grain surfaces. In the midplane, the major ice-phase sulfur carrier is OCS, accounting for $\sim 60$\% of all volatile sulfur below $z/r \lesssim 0.25$. Most of the remaining volatile sulfur in the midplane is locked into CS ice. In the disk atmosphere, the major ice-phase reservoirs are SO$_2$ and SiS. Quantifying the presence of such ices in disks is not straightforward, but we may infer the likely reservoirs through observations of comets, which represent pristine remnants of the primordial solar disk. Detections of OCS from comets are relatively common, with measured abundances ranging from $\sim 0.02 - 0.5$\% relative to water \citep[e.g.][]{dello_russo_1998, le_roy_2015}. OCS ice was also recently tentatively detected in the disk around HH 48 NE with JWST \citep{sturm_2023}. Conversely, SO$_2$ has only two cometary detections, but the inferred abundances are similar to that of OCS \citep{bockelee_morvan_2000, le_roy_2015}. The apparent lack of SO$_2$ detections may be tied to the relative ease with which it is photodissociated, with the photodissociation product SO being much more widely detected \citep[e.g][]{bockelee_morvan_2000,biver_2006, biver_2015, le_roy_2015}. Ice reservoirs may also be inferred from gas-phase emission in disks. Gas-phase SO$_2$ detected towards the Oph~IRS~48, for example, is thought to originate from sublimation of ices at the edge of a dust cavity, with the bulk of the ice reservoir coincident with millimeter dust grains \citep{Booth_2021}. The  presence of SiS ice in our model is more challenging to explain. Silicon is typically considered to be highly depleted in disks, with chemical models using abundances of Si/H $\sim 10^{-8}$, based on low metal abundances inferred from studies of interstellar cloud chemistry \citep{graedel_1982}. Our modelling of HD~100546, however, constrains the total elemental gas-phase silicon abundance in this disk to Si/H $= 5 \times 10^{-7}$, using upper limits of the HCO$^+$ 4-3 disk-integrated line flux. This constraint is due to the impact that electrons from ionized metals (Si, Mg, Fe) have on the disk ionization structure, with high metal abundances leading to suppression of HCO$^+$ \citep{fromang_2002}. Further observations and modelling will be required to better constrain the silicon content, but if the abundance is considerably lower than our modelled value then the constraint from SiS is loosened and a significant amount of additional elemental sulfur may be available for incorporation into other species.

Within the outer dust ring between $r\sim 150-230$ au, molecular ices are dominant at the midplane. OCS and CS are still the major carriers, although H$_2$S ices are also prominant in a narrow region, with abundances that fall off rapidly with height. Further up in the disk atmosphere, OCS ices persist, but CS ice abundances fall off and SiS ices become abundant above $z/r\sim 0.1$. Here, the low densities allow exteral FUV radiation to penetrate more deeply and locally increase the temperatures such that a small amount of sulfur-bearing molecules can exist in the gas-phase. These are predominantly in the form of HS, SO, and CS, although combined account for only $\sim 0.2$\% of sulfur-bearing volatiles.

In summary, our results provide new constraints on the major volatile sulfur carriers in protoplanetary disks. The identification of OCS as a major ice-phase reservoir highlights the need for more high-resolution observations of sulfur-bearing species. Conversely, the low relative abundance of gas-phase H$_2$S in our models places doubt on the consensus that it is the major volatile sulfur carrier in planet-forming environments. The H$_2$S ice reservoir is also small, by comparison to the other sulfur-bearing molecular ices. To assess the impact of initial conditions on the H$_2$S abundance, we ran a second grid of models in which all sulfur was initialized in the form of H$_2$S ice, rather than atomic gas-phase S. We found no considerable difference in the abundances obtained from these models, compared to the fiducial models presented in section \ref{modelling_results}, suggesting that the sulfur chemistry reaches a steady state within the 5 Myr chemical timescale. One possible explanation for the low H$_2$S abundance and large OCS ice reservoir is the lack of grain chemistry in our chemical network. Experimental studies have shown that reactions of solid OCS with atomic hydrogen can produce OCS-H radicals, which further react with H atoms to produce H$_2$S and CO \citep{nguyen_2021}:
\begin{gather}
    \text{OCS + H} \rightarrow \text{OCS-H}
    \label{eq:ocs_1}\\
    \text{OCS-H + H} \rightarrow \text{H}_2 \text{S + CO}
    \label{eq:ocs_2}
\end{gather}
Solid H$_2$S can then be released back into the gas phase via chemical desorption, as described above. Experimental studies also show an efficient depletion of H$_2$S due to H-atom abstraction reactions forming HS, which could associate with other radicals/molecules to form S-bearing species or chemically desorb into the gas phase \citep{oba_2018, santos_2023}. However, we note that our findings on the importance of OCS and relatively low abundance of H$_{2}$S in the HD\,100546 disk mirror those from early solid-state transmission spectroscopy of protostellar ices using JWST, which reach the same conclusion \citep{mcclure_2023}. Such results may be indicative of high levels of UV photoprocessing, which convert a significant amount of H$_2$S ice to OCS \citep{jimenez_escobar_2014, chen_2015}.

\subsection{Implications for planets}
\label{discussion:4_planets}

Elemental abundance ratios in disks are promising tracers of planetary formation and migration pathways. The C/O ratio has received particular attention in studies of both disks and exoplanet atmospheres due to the high cosmic abundances of both carbon and oxygen, and the wealth of associated molecular carriers. Simple models relate C/O measured in exoplanet atmospheres to potential formation locations in disks based on the location of volatile molecular snowlines. In this approach, the gas-phase C/O is a step function which increases with distance from the host star \citep{oberg2011}. Recent studies have expanded on this idea, showing that additional mechanisms can lead to both radial and azimuthal C/O variations at small spatial scales, which adds further complexity \citep{cleeves2018, zhang_2019, miotello_2019, maps_7_bosman2021}. While the C/O ratio remains a powerful diagnostic tool, its utility as a tracer of planetary formation processes is limited, especially in cases where planets acquire the bulk of their carbon and oxygen beyond the CO$_2$ snowline. In recent years, the use of additional elemental abundance ratios has been explored to further constrain disk compositions and break degeneracies associated with the C/O ratio. \citet{turrini_2021}, for example, showed that the C/N and N/O ratios can be used to constrain the extent of planetary migration, while the S/N ratio can be used as a reliable tracer of heavy element enrichment. Further, \citet{hobbs_2022} used the C/H, O/H, and N/H ratios to predict connections between the atmospheric molecular abundances of Hot Jupiters and their formation pathways. In this section, we explore the utility of the S/H ratio and its impact on planets in the HD~100546 disk, as well as its wider application to disks in general.

In HD~100546, multiple observations provide evidence for at least two giant planetary candidates within the disk; planet `b' at $r \sim 55$ au, and planet `c' within the inner cavity at $r \sim 13$ au. How might the composition of these planets atmospheres vary, based on the radial gas-phase S/H variations implied by our models? As we have shown, a region of elevated \shgas \ in the outer disk appears broadly coincident with the outer millimeter dust ring, while another region of elevated \shgas\ resides directly outside of the inner dust ring. The dust rings themselves are likely the products of planetary formation processes, and the interplay between the dust structure, gas composition, and materials accreted by planets will be complex and ever-evolving. It is important to highlight the fact that the vast majority of elemental sulfur ($>94$\%) is locked up into refractory materials, and the sulfur content of an exoplanet atmosphere will not only depend on the composition of the accreted gas, but the amount of sulfur returned to the gas phase from accreted solids. Gas giant envelopes can become enriched in this manner through a number of mechanisms, including dissolution of pebbles/planetesimals during accretion, and core erosion/mixing \citep{hueso_guillot_2003, estrada_2009}. Some level of enrichment is to be expected, but placing precise constraints is challenging since dissolution and mixing are not well understood \citep{leconte_chabrier_2013, nettelmann_2015, moll_2017}. Studies of elemental enrichment in Jupiter's atmosphere, for example, suggest that $> 50$\% of the core has become dissolved in the envelope \citep{oberg_wordsworth_2019}. Such high levels of mixing may not be representative of the general exoplanet population, however, since it has been proposed that the distribution of heavy elements in Jupiter's inner envelope could only come about through a collision which shattered Jupiter's primordial core \citep{liu_2019}. Uncertainties on the levels of enrichment through core dissolution/mixing are particularly problematic for interpreting observations of sulfur-bearing species in exoplanet atmospheres, since almost all sulfur is tied up in refractory species. \looseness=-1

Previous studies which propose the S/N ratio as direct tracer of planetesimal accretion, rely on the fact that most of the sulfur is incorporated into solids, while nearly all nitrogen is in gaseous N$_2$ \citep{turrini_2021}. The S/H ratio is analagous to the S/N ratio in this sense. The joint evaluation of S/N with independent measurements of metallicity allow further constraints to be placed on migration scenarios, with in-situ formation characterised by stellar metallicity and stellar S/N, and extensive migration characterised by super-steller metallicity and stellar S/N \citep{pacetti_2022}. In practice, observations and modelling of sulfur-bearing species in exoplanet atmospheres will need to consider the separate contributions from  directly-accreted gas, and solids returned to the gas-phase, in order to use S/N or S/H as reliable tracers of these mechanisms.


\section{Conclusions}

In this work, we presented new data from ALMA and APEX relating to volatile sulfur-bearing species in the HD100546 protoplanetary disk. We used a highly refined and disk-specific physical-chemical models to infer the total elemental sulfur content, and predict which species are the major carriers of volatile sulfur in the disk. Our main conclusions are:

1. We observed transitions of seven sulfur-bearing species towards HD~100546 with ALMA, in addition to an observation of H$_2$S at 168.762 GHz with APEX. All species were undetected. We derived upper limits on the disk-integrated fluxes for all transitions.

2. We used the physical-chemical disk modelling code \textsc{dali} to determine the volatile sulfur abundance. Using simple models with a single sulfur abundance across the entire disk, we find volatile sulfur to be heavily depleted from the cosmic value by at least two orders of magnitude, S/H $\lesssim 5 \times 10^{-8}$, constrained by these upper limits and the previous detections of SO and CS

3. Comparing our models to the \sostack radial intensity profile from \citet{booth_2023a}, we find that a single sulfur abundance across the entire disk does not adequately reproduce the observations. We used an MCMC approach to construct a composite model, finding that the sulfur abundance varies radially throughout the disk by at least three orders of magnitude.

4. Radial variations in the sulfur abundance broadly coincide with the location of prominent millimeter dust rings and gaps. Models indicate that many of the sulfur-bearing molecules are particularly abundant inside the inner dust ring ($r\sim 22-40$ au), and coincident with the outer dust ring ($r \sim 150-230$ au). The total gas-phase sulfur abundance is high within the dust-depleted inner cavity, with most sulfur in the form of atomic S or S$^+$. SO and HS are most abundant molecular carriers with inner cavity.

5. We determined the major volatile gas-phase sulfur carriers to include OCS, H$_2$CS, and CS. We also infer the presence of a substantial OCS ice reservoir in the outer disk, alongside other molecular ices at lower abundance. H$_2$S is found to have low abundances in both the gas- and ice-phases, in contrast to predictions from many previous astrochemical models.

6. The total amount of sulfur held in volatiles ranges from $\sim 0.01$ to $5.3$\%, dependent on radius. The use of elemental ratios S/H and S/N as tracers of planetary formation is complicated by the uncertainties on the split between volatiles and refractories, and the level at which accreting refractories are processed into planetary atmospheres.

\section*{Acknowledgements}

L.K. acknowledges funding via a Science and Technology Facilities Council (STFC) studentship. K.J.C. is grateful for the Dutch Research Council (NWO) support via a VENI fellowship (VI.Veni.212.296). L.I.C. acknowledges support from NASA ATP 80NSSC20K0529, NSF AST-2205698, the David and Lucille Packard Foundation, the Research Corporation for Scientific Advancement Cottrell Scholar Award, and Heising-Simons Grant 2022-3995. M.N.D. acknowledges the Swiss National Science Foundation (SNSF) Ambizione grant number 180079, the Center for Space and Habitability (CSH) Fellowship, and the IAU Gruber Foundation Fellowship.

\section*{Data availability}

The data presented here are from the ALMA Cycle 5 programme 2017.1.00885.S (principal investigator M. Kama). The raw data are publicly available from the ALMA archive. The reduced data and final imaging products are available upon reasonable request from the corresponding author. The ALMA data were reduced using CASA version 5.6.1-8, which is available at https://casa.nrao.edu/.



\bibliographystyle{mnras}

 \newcommand{\noop}[1]{}



\appendix

\clearpage

\section{ALMA continuum observation} \label{appendix:continuum}

\begin{figure}
\centering
\includegraphics[clip=,width=1.0\linewidth]{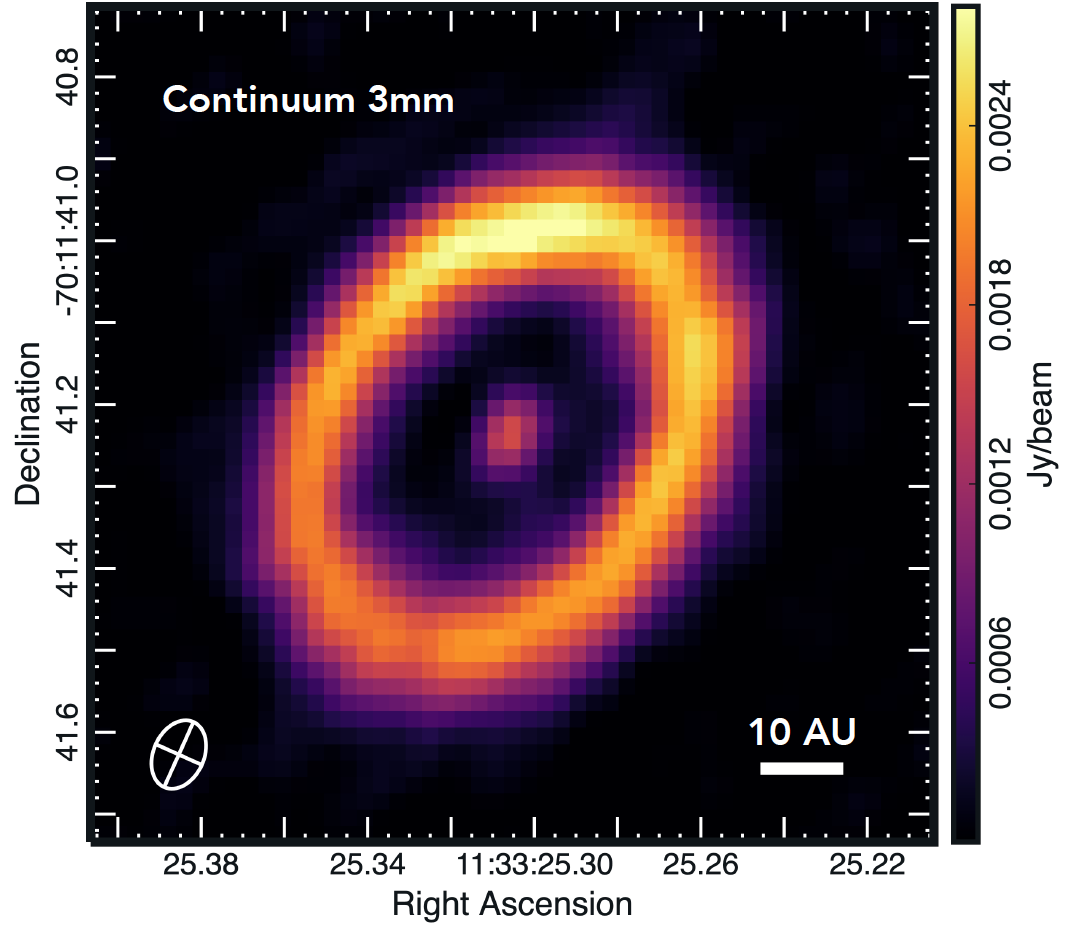}
\caption{ALMA 3mm continuum observation. The inner dust ring ($r\sim 22$ to $40$ au) is well-resolved, and the small inner dust disk is detected ($r \sim 1$ au). Beam size indicated by the white ellipse in the lower left.}
\label{fig_spectra}
\end{figure}

\clearpage

\section{\textsc{dali} fiducial model parameters and updates} \label{appendix:model_params_updates}

We improve the fit to the dust continuum emission by simultaneously fitting 870$\mu$m continuum radial profile \citep{fedele_2021} and 3mm continuum radial profile (this work). We find that the combined data are best fit with a central cavity that extends to $\sim 22$ au, compared to 13 au used in our previous study. We fit the ring-like structure of the outer disk by fully depleting the model of large dust grains between 40-150 au.

To ensure that the SED remains well-fit, we run a large grid of $\sim 5000$ models using the updated dust structure. We vary the parameters $r_\textrm{gap}, \;\psi, \;h_\textrm{c}, \;\Delta_{g/d}, \;\delta_\textrm{dust}, \;\chi, \;f$, and $\Sigma_\textrm{c}$, finding that the new model is best-fit by increasing the settling parameter $f$ from 0.85 to 0.95, and the flaring index $\psi$ from 0.20 to 0.25. The critical surface density, $\Sigma_\text{c}$ is slightly increased from 82.75 g cm$^{-2}$ to 87.25 g cm$^{-2}$, such that the total dust mass remains roughly the same as our original model, accounting for the mass removed by extending the cavity and introducing an outer dust gap. All other physical model parameters remain unchanged.

We explore the  impact of altering the dust structure on the chemical abundances by running a grid of models in which [C/H] and [O/H] are varied between $2 \times 10^{-6}$ and $2 \times 10^{-4}$. For each case, we consider variations in the level of gas depletion in the central cavity, ranging from $\delta_\text{gas} = 10^{-5}$ to $1$. As done previously in \citet{keyte_2023}, we fit to the CO ladder, a wide range of other disk-integrated line fluxes, and various spectral line and radial intensity profiles (see that work for a full list of the data). As before, we find that the model is best fit with a C/O ratio of 0.5, with [C/H] slightly increased from $1 \times 10^{-5}$ to $1.5 \times 10^{-5}$, and [O/H] slightly increased from $2 \times 10^{-5}$ to $3 \times 10^{-5}$. Gas within the central cavity is found to be severely depleted, with a best-fit value of $\delta_\text{gas} = 10^{-5}$.

Fiducial model parameters are listed in Table \ref{table:modelparameters} and initial atomic abundances are listed in Table \ref{table:initial_abundances}.

\begin{table*}
\caption{Fiducial HD~100546 disk model parameters.}             
\label{table:modelparameters}      
\centering
\begin{tabular}{l l l}     %
\hline\hline       
                      
Parameter & Description & Fiducial\\ 
\hline                    
   \rsub                & Sublimation radius                         & 0.25 au    \\
   \rgap                & Inner disk size                            & 1.0 au     \\
   \rcav                & Cavity radius                              & 22 au      \\
   $R_\text{out}$       & Disk outer radius                          & 1000 au    \\
   $R_c$                & Critical radius for surface density        & 50 au      \\
   \deltagas            & Gas depletion factor inside cavity         & $10^{-5}$  \\
   \deltadust           & Dust depletion factor inside cavity        & $10^{-4}$  \\
   $\gamma$             & Power law index of surface density profile & 1.0        \\
   $\chi$               & Dust settling parameter                    & 0.2        \\
   $f$                  & Large-to-small dust mixing parameter       & 0.95       \\
   $\Sigma_c$           & $\Sigma_\text{gas}$ at $R_c$               & 87.25 g cm$^{-2}$  \\
   $h_c$                & Scale height at $R_c$                      & 0.10       \\
   $\psi$               & Power law index of scale height            & 0.25       \\
   \gasdust             & Gas-to-dust mass ratio                     & 100        \\
   $L_*$                & Stellar luminosity                         & $36\; L_\odot$        \\
   $L_X$                & Stellar X-ray luminosity                   & $7.94 \times 10^{28} \text{ erg s}^{-1}$    \\
   $T_X$                & X-ray plasma temperature                   & $7.0 \times 10^{7}$ K     \\
   $\zeta_\text{cr}$    & Cosmic ray ionization rate                 & $3.0 \times 10^{-17}$ s$^{-1}$  \\
   $M_\text{gas}$       & Disk gas mass                              & $9.89 \times 10^{-2}$ \msun   \\
   $M_\text{dust}$      & Disk dust mass                             & $1.06 \times 10^{-3}$ \msun   \\
   $t_\text{chem}$      & Timescale for time-dependent chemistry     & \text{5 Myr} \\
\hline                  
\end{tabular}
\end{table*}

\begin{table*}
\caption{Initial elemental abundances.}             
\label{table:initial_abundances}      
\centering
\begin{tabular}{l l}     %
\hline\hline       
                      
Species & $n_i/n_H$ \\ 
\hline                    
   H  & 1.0  \\
   He & $9.0 \times 10^{-2} $ \\
   C  & $1.5 \times 10^{-5} $ \\
   N  & $6.2 \times 10^{-5} $ \\
   O  & $3.0 \times 10^{-5} $ \\
   S  & $1.0 \times 10^{-8} $ \\
   Mg & $5.0 \times 10^{-7} $ \\
   Si & $5.0 \times 10^{-7} $ \\
   Fe & $5.0 \times 10^{-7} $ \\
\hline            
\end{tabular}
\end{table*}

\section{Abundance maps}\label{appendix:abundance_maps}

\begin{figure*}
\centering
\includegraphics[clip=,width=1.0\linewidth]{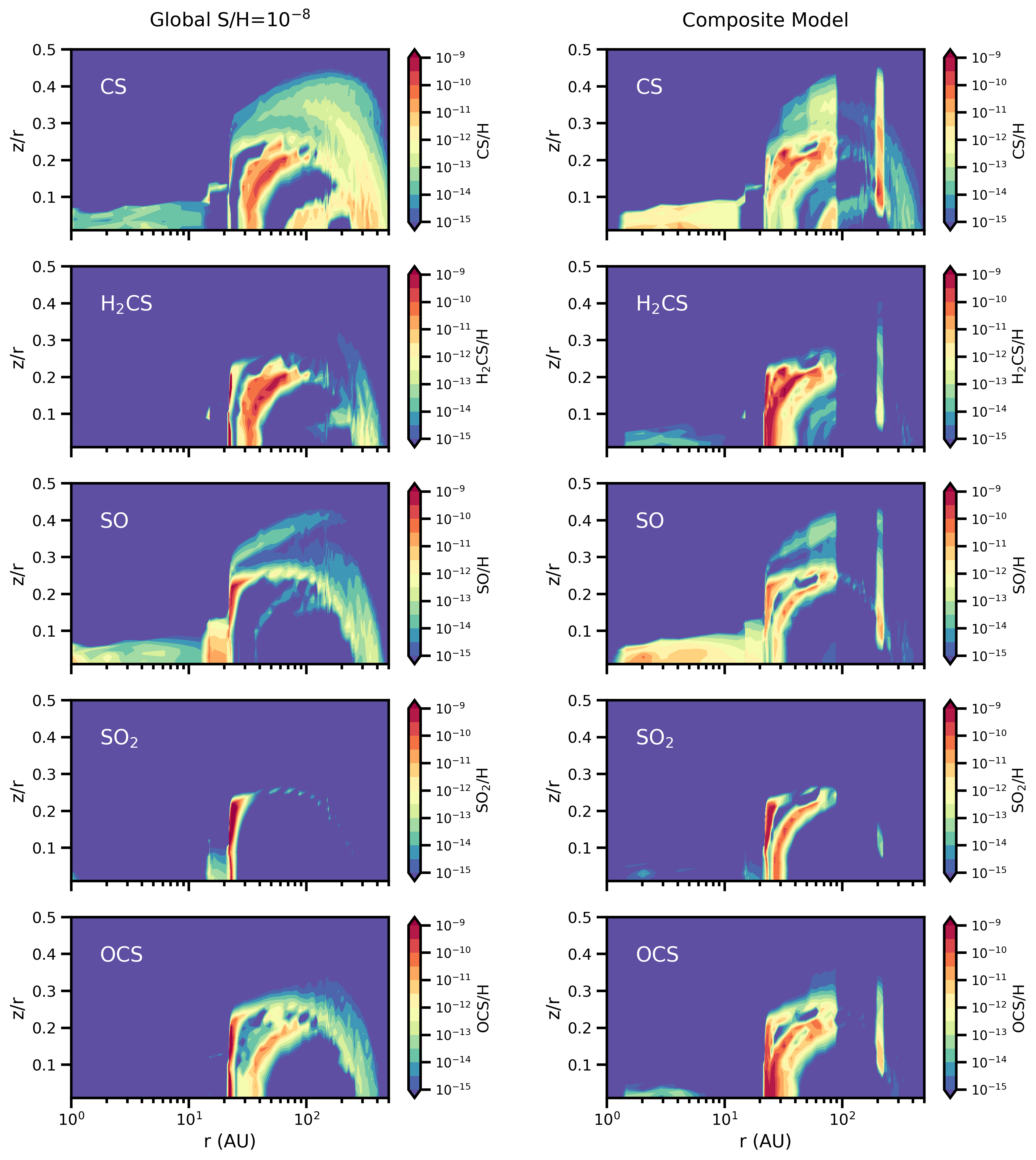}
\caption{Abundance maps for the carbon- and oxygen-bearing sulfur species in models. \emph{Left column: }Model using a single sulfur abundance across the entire disk (S/H=$10^{-8}$). \emph{Right column: }Composite model using a radially varying sulfur abundance profile.}
\label{fig_abundancemaps_carbonoxygen}
\end{figure*}

\begin{figure*}
\centering
\includegraphics[clip=,width=1.0\linewidth]{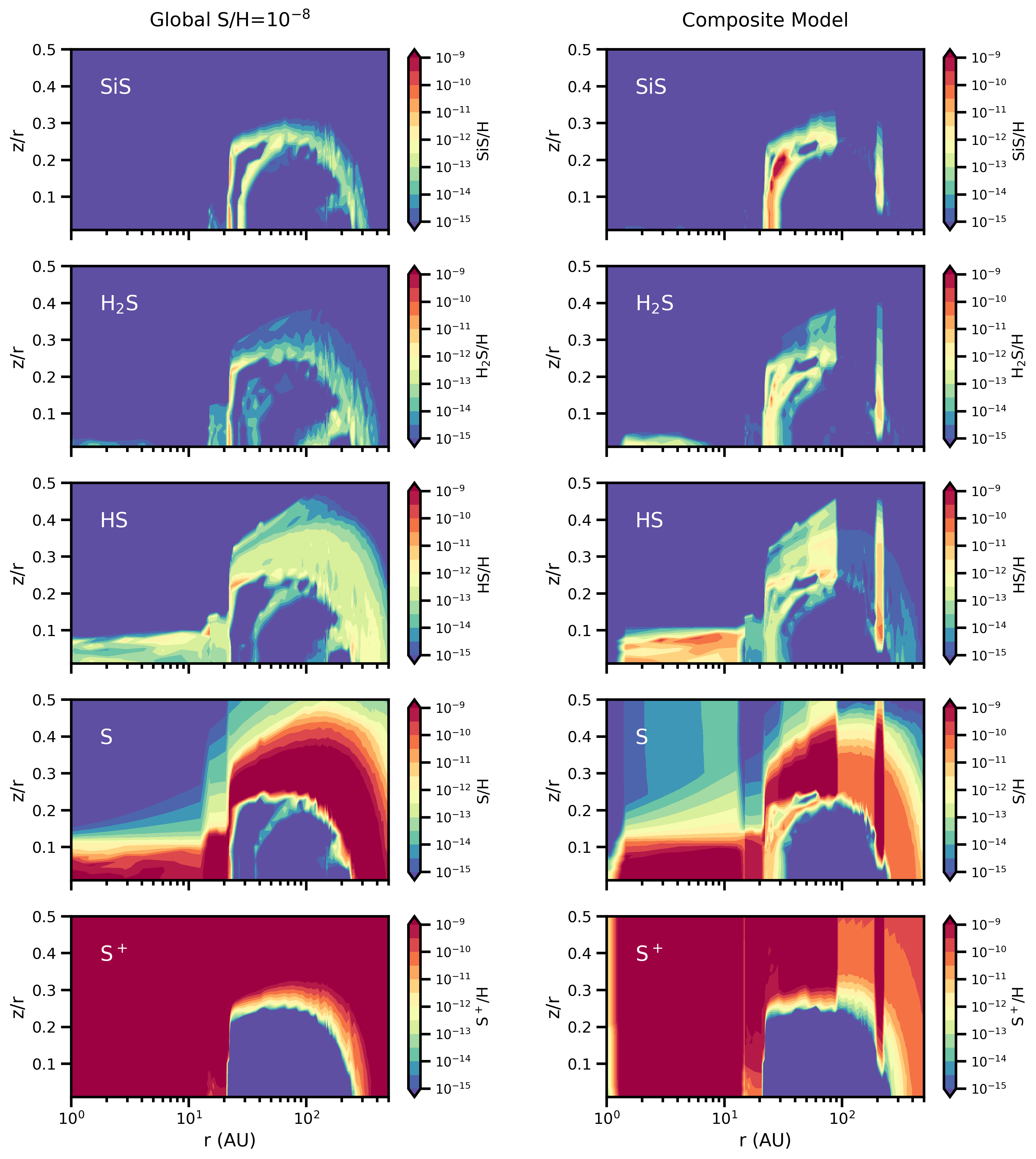}
\caption{Abundance maps for the sulfur species containing no carbon or oxygen in our models. \emph{Left column: }Model using a single sulfur abundance across the entire disk (S/H=$10^{-8}$). \emph{Right column: }Composite model using a radially varying sulfur abundance profile.}
\label{fig_abundancemaps_other}
\end{figure*}

\begin{figure*}
\centering
\includegraphics[clip=,width=1.0\linewidth]{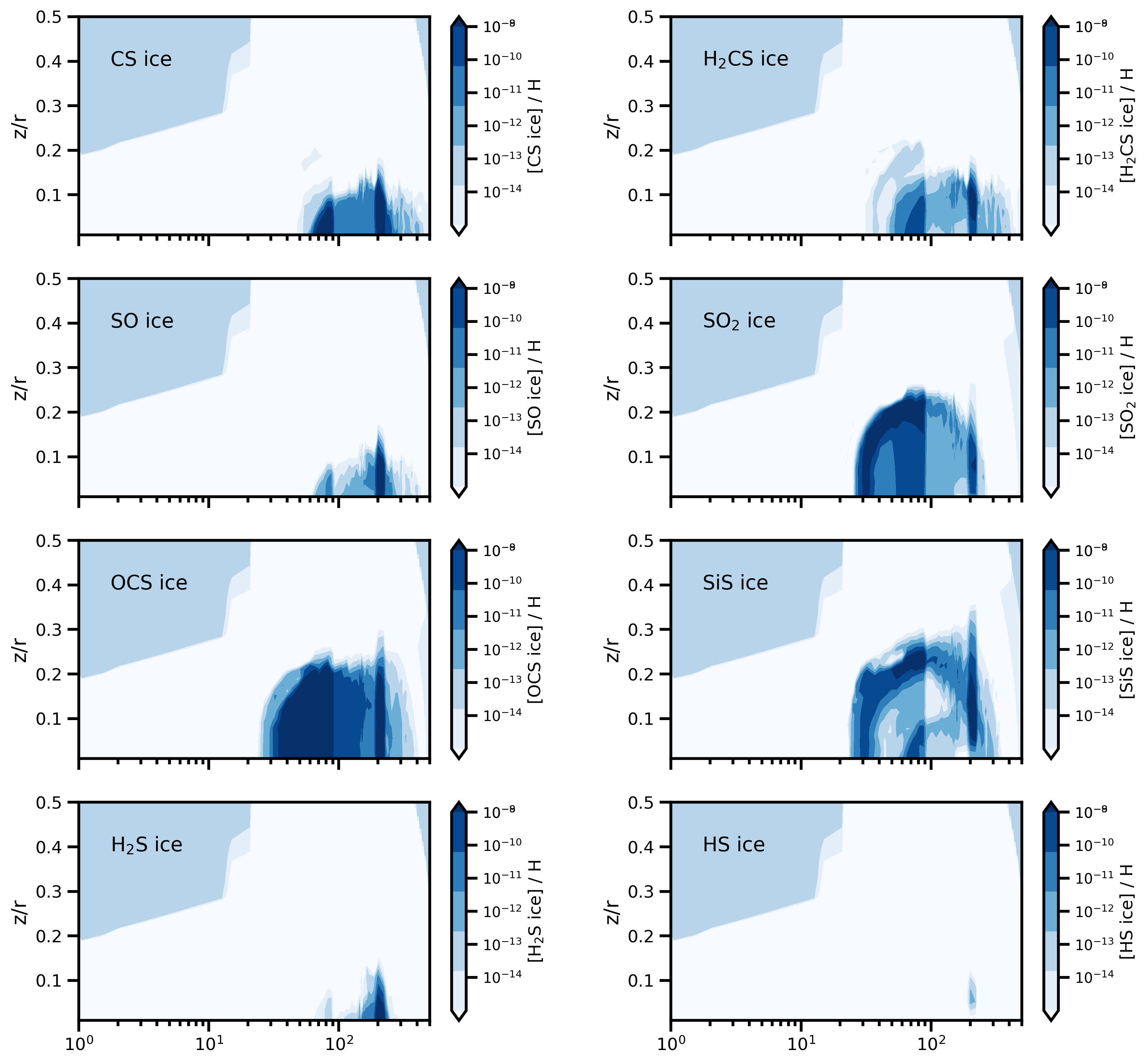}
\caption{Abundance maps for sulfur-bearing ices in our composite model.}
\label{fig_abundancemaps_ices}
\end{figure*}

We present 2D abundance maps for 10 key molecular sulfur-bearing species extracted from our models. Figure \ref{fig_abundancemaps_carbonoxygen} shows abundance maps for sulfur-bearing species which contain carbon or oxygen, and Figure \ref{fig_abundancemaps_other} shows abundance maps for sulfur-bearing species containing neither carbon or oxygen. In each case, the left column shows abundances extracted from our model using a single initial sulfur abundance (S/H $=10^{-8}$) and the right column shows abundances extracted from the composite model. Figure \ref{fig_abundancemaps_ices} shows abundances for all species in the ice-phase, extracted from the composite model.

\section{Radial intensity profile MCMC fitting} \label{appendix: MCMC}

\begin{figure*}
\centering
\includegraphics[clip=,width=1.0\linewidth]{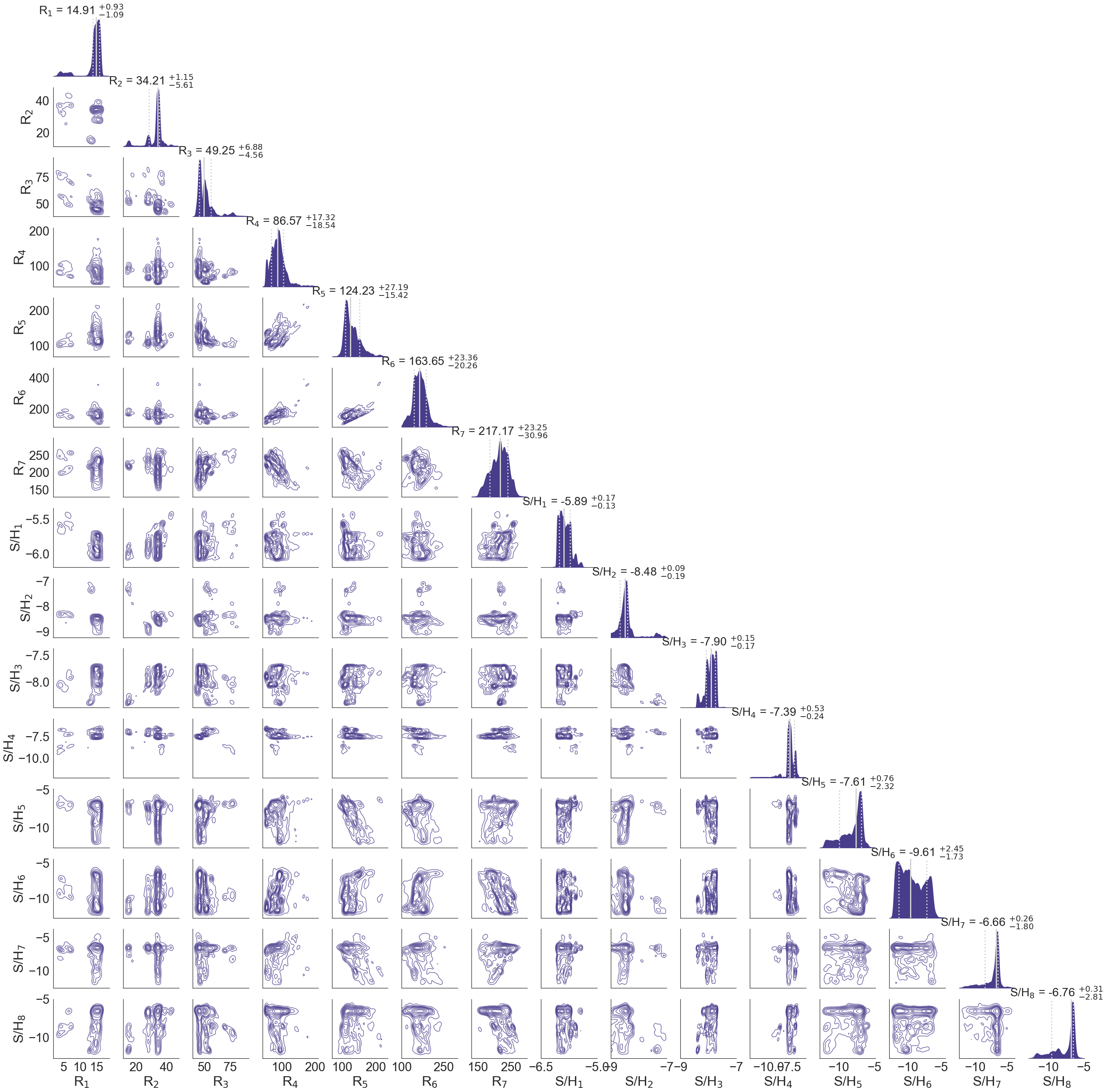}
\caption{Posterior distribution for parameters used to simultaneously fit the \sostack 
 and CS 7-6 radial intensity profiles. Parameters labelled R$_n$ are the radial locations where different models are spliced together (in au). Parameters labelled S/H$_n$ are the log10 sulfur abundances of the models fitted to each radial region.}
\label{fig_corner}
\end{figure*}

\begin{figure*}
\centering
\includegraphics[clip=,width=1\linewidth]{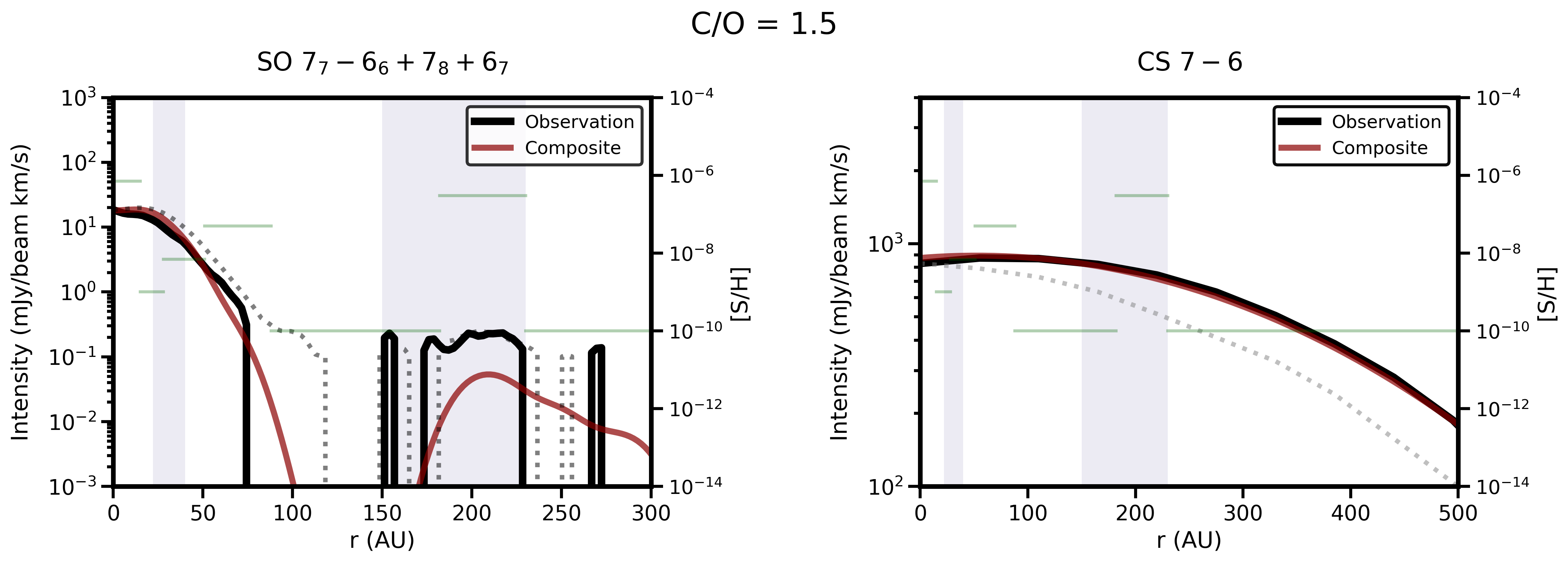}
\caption{Radial intensity profiles extracted from deprojected integrated intensity maps, azimuthally averaged in the region of the disk where C/O\;=\;1.5. Observations (black) and composite models (red). The models used in this region are initialised with abundances extracted from the C/O\;=\;0.5 model. Oxygen is then removed such that gas-phase C/0=1.5, and the chemistry is evolved for 5 years, following the procedure outlined in \citet{keyte_2023} Dotted grey lines show the observed radial intensity profiles in the region where C/O=0.5, for comparison.}
\label{fig_radial_co1p5}
\end{figure*}

Figure \ref{fig_corner} shows the posterior distribution for our MCMC fitting of the composite model. Figure \ref{fig_radial_co1p5} shows the results from fitting our composite model to the region in which C/O=1.5.

\clearpage

\section{Comparison with STARCHEM chemical modelling code}\label{appendix:starchem_comparison}


To assess the validity of the sulfur chemistry in our \textsc{dali} models, we ran a set of comparison models using the \textsc{starchem} chemical modelling code (\citealp{rawlings_2022}, Rawlings, Keto \& Caselli, submitted), which incorporates the full RATE12 dataset (467 species and 5950 reactions; \citealp{mcelroy_umist12}). Our goal was not to repeat the comprehensive modelling of the entire disk, but instead to perform a more in-depth examination of the chemistry, with a specific emphasis on validating the presence of OCS as a significant sulfur carrier in the gas phase. Such a comparison offers valuable insights into chemical network dependencies and potential consequences of omitting crucial species or reaction pathways from the \textsc{dali} models.

To this end, we compared the chemistry at two vertical locations in the disk inside the inner dust ring at 22 au; the midplane (z/r=0) and disk atmosphere (z/r=0.2). These regions were selected as points at which the \textsc{dali} model predicts OCS abundances to be relatively high in the gas phase (OCS/H $\sim 10^{-9}$ in both locations). The inputs to \textsc{starchem} were gas and dust densities/temperatures, cosmic-ray and X-ray ionization rates, and the UV radiation field, each of which were taken from the relevant gridpoint locations in the \textsc{dali} model. The chemistry was then initialised with elemental abundances following the \textsc{dali} model (Table \ref{table:initial_abundances}), and evolved for 5 Myr. 




The outputs from \textsc{starchem} broadly agree with our \textsc{dali} model; we find OCS to be one of the major gas-phase sulfur carriers, with an abundance of OCS/H $\sim 2 \times 10^{-10}$ at both locations. In the \textsc{starchem} model, OCS formation is dominated by:
\begin{gather}
    \text{CO + S}^- \rightarrow \text{OCS + e}^-
    \label{eq:ocs_starchem1}
\end{gather}
with some contribution from:
\begin{gather}
    \text{CO + HS} \rightarrow \text{OCS + H}
    \label{eq:ocs_starchem2}\\
    \text{OH + CS} \rightarrow \text{OCS + H}
    \label{eq:ocs_starchem3}
\end{gather}
Our \textsc{dali} model does not include S$^-$, and OCS is instead formed primarily by:
\begin{gather}
    \text{CO + S} \rightarrow \text{OCS}
    \label{eq:ocs_dali1}
\end{gather}
The main OCS destruction pathways in both models are cosmic-ray and X-ray induced photodissociation and photoionisation.

In additional to OCS, the \textsc{starchem} model predicted the other major sulfur-carriers at this location to be H$_2$CS and SO$_2$. The \textsc{starchem} H$_2$CS abundances of $2.5 \times 10^{-10}$ ($z/r=0$) and $2.0 \times 10^{-10}$ ($z/r=0.2$) are in agreement with the outputs of our \textsc{dali} model within an order of magnitude. The SO$_2$ abundances are slightly more divergent. The miplane abundance in the \textsc{starchem} model is a factor of $\sim 10$ higher than the \textsc{dali} model, with formation facilitated via the route:
\begin{gather}
    \text{HS + S} \rightarrow \text{S}_2 \text{ + H}
    \label{eq:so2_starchem1}\\
    \text{S}_2 \text{+ O} \rightarrow \text{SO + S}
    \label{eq:so2_starchem2}
\end{gather}
and
\begin{gather}
    \text{OH + S} \rightarrow \text{SO} \text{ + H}
    \label{eq:so2_starchem3}\\
    \text{SO + OH} \rightarrow \text{SO}_2 \text{ + H}
    \label{eq:so2_starchem4}
\end{gather}
or
\begin{gather}
    \text{SO + O} \rightarrow \text{SO}_2
    \label{eq:so2_starchem5}
\end{gather}
The higher SO$_2$ abundance in the \textsc{starchem} model is likely due to the lack of S$_2$ in the \textsc{dali} network, with SO$_2$ formed predominantly only through reactions \ref{eq:so2_starchem4} and \ref{eq:so2_starchem5} in the \textsc{dali} model.

In the disk atmosphere, the SO$_2$ abundance in the \textsc{starchem} model falls off dramatically (SO$_2$/H $=8\times 10^{-18}$), with S$_2$ emerging as the main volatile sulfur carrier (S$_2$/H $=3\times 10^{-9}$). A similar drop in the SO$_2$ abundance is seen in the \textsc{dali} model at a slightly higher vertical location ($z/r\sim0.25$). In both models, the main SO$_2$ destruction pathways are once again cosmic-ray and X-ray induced photodissociation and photoionisation.

These findings offer valuable perspectives on sulfur chemistry, but it's important to acknowledge that there are likely to be important processes not captured in either model. For instance, although the chemical network in the \textsc{starchem} model is significantly more complex, it does not include gas-grain reactions, three-body reactions, and PAHs. The absence of gas-grain reactions could have significant ramifications for OCS-related chemistry, as laboratory experiments have demonstrated it is readily formation within icy grain mantles and can undergo significant processing \citep{ferrante_2008, nguyen_2021}. More work is needed to develop a deeper understanding of sulfur chemistry in planet-forming environments.

\bsp	
\label{lastpage}
\end{document}